\documentclass[]{spie}  

\usepackage{amsmath,amsfonts,amssymb}
\usepackage{graphicx}
\usepackage[colorlinks=true, allcolors=blue]{hyperref}
\usepackage{textcmds}
\usepackage{aas_macros}
\usepackage[utf8]{inputenc}

\title{Evaluation of scientific CMOS sensors for sky survey applications}

\author[a]{Sergey Karpov}
\author[a]{Armelle Bajat}
\author[a]{Asen Christov}
\author[a]{Michael Prouza}
\author[b,c]{Grigory Beskin}

\affil[a]{CEICO, Institute of Physics, Czech Academy of Sciences, Prague, Czech Republic}
\affil[b]{Special Astrophysical Observatory, Nizhniy Arkhys, Russia}
\affil[c]{Kazan Federal University, Kazan, Russia}

\authorinfo{Further author information: (Send correspondence to S.K.)\\S.K.: E-mail: karpov@fzu.cz}

\begin{document}
\maketitle

\begin{abstract}
  Scientific CMOS image sensors are a modern alternative for a typical CCD detectors, as they offer both low read-out noise, large sensitive area, and high frame rates. All these makes them promising devices for a modern wide-field sky surveys. However, the peculiarities of CMOS technology have to be properly taken into account when analyzing the data. In order to characterize these, we performed an extensive laboratory testing of two Andor cameras based on sCMOS chips -- Andor Neo and Andor Marana. Here we report its results, especially on the temporal stability, linearity and image persistence. We also present the results of an on-sky testing of these sensors connected to a wide-field lenses, and discuss its applications for an astronomical sky surveys.
\end{abstract}

\keywords{Calibration, CMOS sensors, Sky surveys}

\section{Introduction}\label{sec_intro}

Sky survey applications require large format image sensors with high quantum efficiency, low read-out noise, fast read-out and a good inter- and cross-pixel stability and linearity. Charge-Coupled Devices (CCDs), typically employed for such tasks, lack only the read-out speeds, which significantly lowers their performance for detecting and characterizing rapidly varying or moving celestial objects. On the other hand, Complementary Metal–Oxide–Semiconductor (CMOS) imaging sensors, widely used on the consumer market, typically displays levels of read-out noise unacceptably large for precise astronomical tasks (tens of electrons), as well as significantly worse uniformity and stability than CCDs.

However, recent development in the low-noise CMOS architectures (see e.g. \cite{spie_scmos}) allowed to design and create a market-ready large-format (2560x2160 6.5$\mu$m pixels) CMOS chips with read-out noise as low as 1-2 electrons, on par with best CCDs \cite{spie_neo} -- so-called ``scientific CMOS'' (sCMOS) chips. Like standard CMOS sensors (and unlike CCDs), they did not perform any charge transfer between adjacent pixels, employing instead individual column-level amplifiers with parallel read-out and dual 11-bit analog-to-digital converters (ADCs) operating in low-gain and high-gain mode, correspondingly, and an on-board field-programmable gate array (FPGA) logic scheme that reconstructs a traditional 16-bit reading for every pixel from two 11-bit ones.

The cameras based on this original sCMOS chip, CIS2051 (later rebranded as CIS2521) by Fairchild Imaging, have been available since 2009. Andor Neo is one of such cameras and is currently widely used in astronomical applications, especially for tasks that require high frame rates like satellite tracking \cite{schildknecht_2013}, fast photometry \cite{qui_2013} or rapid optical transients detection \cite{karpov_2019}. Detailed characterization of this camera presented in these works displays its high performance and overall good quality of delivered data products, with the outlined problems related mostly to the non-linearity at and above the amplifier transition region around 1500 ADU and a low full well depth of about only 20k electrons.

Scientific CMOS sensors with larger well depth, up to 120k electrons \cite{spie_400}, and a back-illuminated options having significantly better quantum efficiency (up to 95\%) have been released later by a GPixel. Andor Marana \cite{andor_marana} is a camera built around such back-illuminated chip, GSense400BSI. Its parameters (see Table~\ref{tab_marana}) look especially well suited for a wide range of astrophysical tasks -- detection and study of rapid optical transients \cite{karpov_2010, karpov_2019}, space debris tracking \cite{karpov_2016} or observations of faint meteors \cite{karpov_meteors_2019}. Due to large frame format, absence of microlens raster on top of the chip, good quantum efficiency and fast read-out, such device -- if proven to be stable enough -- may also be a promising detector for a next generation of FRAM atmospheric monitoring telescopes \cite{fram,fram_cta}, which shall perform rapid and precise stellar photometry in a very wide fields of view in order to assess atmospheric variations in real time.

Thus, we decided to perform the laboratory and on-sky characterization of Andor Marana camera, generously provided to us by the manufacturer for testing. We also run some of the tests on Andor Neo cameras we have in order to compare the properties of these two generations of sCMOSes based on different sensors. As both cameras perform quite complex onboard post-processing of acquired images, we will concentrate just on their major user-visible properties, especially the ones important for the tasks of high temporal resolution sky surveys -- i.e. on pixel-level stability and uniformity.

\begin{table}[t]
  \caption{Comparison of parameters of Andor Neo and Andor Marana sCMOS cameras, taken from the specifications from manufacturer website. Actual values of parameters like dark current, readout noise and full well capacity, vary for every individual camera, and have to be measured independently.}
  \label{tab_marana}
  \begin{center}
    \begin{tabular}{|l|c|c|}
      \hline
      \textbf{} & \textbf{Andor Neo 5.5}  & \textbf{Andor Marana 4.2B-11} \\
      \hline
      Chip & CIS2521 & GSense400BSI \\
                & Front illuminated & Back illuminated \\
                & + microlens (f/1.6) raster & \\
      Format & 2560 x 2160 & 2048 x 2048 \\
      Diagonal & 21.8 mm & 31.8 mm \\
      Pixel size  & 6.4 $\mu$m & 11 $\mu$m \\
      Peak QE & 60\% & 95\% \\
      Shutter Mode& Rolling, Global & Rolling \\
      Max FPS & 30 & 48 \\
      Readout noise & 1.0 e$^-$ & 1.6 e$^-$ \\
      Dark current,  e$^-$/pix/s & 0.015 at -30$^\circ$C & 0.7 at -25$^\circ$C \\
      Gain, e$^-$/ADU & 0.67 (rolling) 1.91 (global)& 1.41 \\
      Full well depth  & 30000 e$^-$ & 85000 e$^-$ \\
      I/O Interface & CameraLink & USB3.0 \\
      \hline
    \end{tabular}
  \end{center}
\end{table}

\section{Experimental setup}


For testing the Marana camera, we used an existing experimental equipment available at a laboratory of characterization of optical sensors for astronomical applications at Institute of Physics of Czech Academy of Sciences. 
That included a fully light-isolating dark box, CAMLIN ATLAS 300 monochromator, CAMLIN APOLLO X-600 Xenon lamp, and an integrating sphere mounted directly on the input port of a dark box. A dedicated photodiode coupled with Keithley picoammeter is used to control the intensity of light inside the integrating sphere. The whole system is controlled by a dedicated {\sc CCDLab} software \cite{ccdlab}, which performs real-time monitoring of a system state and stores it to a database for analyzing its evolution, displays it in a user-friendly web interface, and also allows easy scripting control of its operation. The camera itself was controlled by a {\sc FAST} data acquisition software \cite{fast} specifically designed for operating fast frame rate scientific cameras of various types.

As the laboratory environment we used was not dust-free, the camera was equipped with a Nikkor 300 f/2.8 lens, adjusted in such way as to provide illumination of the whole chip with the light from integrating sphere output window. Due to lens vignetting, this resulted in a slightly bell-shaped flat fields. The same lens has been later used for the on-sky testing of the camera photometric performance. For it, the camera with lens were installed on a Software Bisque' Paramount ME mount, controlled with {\sc RTS2} software \cite{rts2}.
The whole setup was installed in the dome of FRAM telescope being tested and commissioned at the same time at the backyard of Institute of Physics of Czech Academy of Sciences in Prague.

For the evaluation of Andor Neo we used one of the cameras of the Mini-MegaTORTORA multi-channel wide-field monitoring system \cite{karpov_2019}, installed in Special Astrophysical Observatory close to Russian 6-m telescope. During the experiments, the cover of the channel containing the camera was closed in order to ensure proper dark frame acquisition, and the built-in channel light source (non-stabilized LED) was used to illuminate inner side of the cover to provide quasi-flat field illumination when necessary. The camera was controlled by the dedicated Mini-MegaTORTORA data acquisition software \cite{beskin_astbull}.

The testing consisted by a scripted set of imaging sequences of various exposures and durations, acquired under different light intensities or in the dark, and with different readout settings. Most of the sequences were acquired with chip temperature set to -30$^\circ$C for Marana, which the camera' Peltier cooler was able to continuously support within the closed area of the dark box with just an air cooling during the whole duration of our experiments, and with -20$^\circ$ for Andor Neo, which is also an optimal sustainable temperature for its setup.
However, for some tests the temperature was also varied. All tests have been performed with a dual amplifier (16-bit) regime, as a most convenient for astronomical applications. For Andor Marana, both global shutter and rolling shutter settings were tested.


\section{Dark current}\label{sec_dark}

\begin{figure}[t]
  \centerline{
    \resizebox*{0.365\columnwidth}{!}{\includegraphics[angle=0]{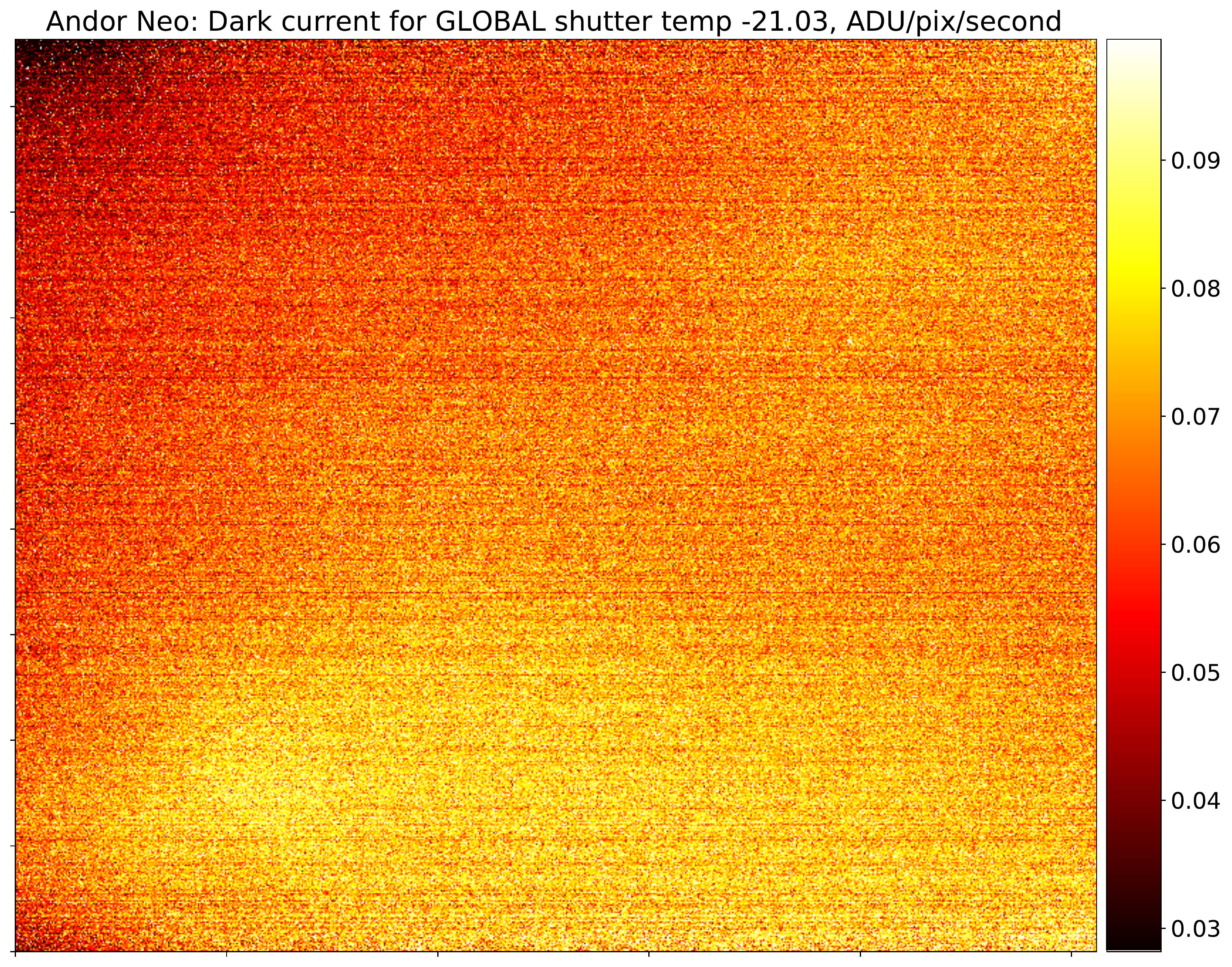}}
    \resizebox*{0.31\columnwidth}{!}{\includegraphics[angle=0]{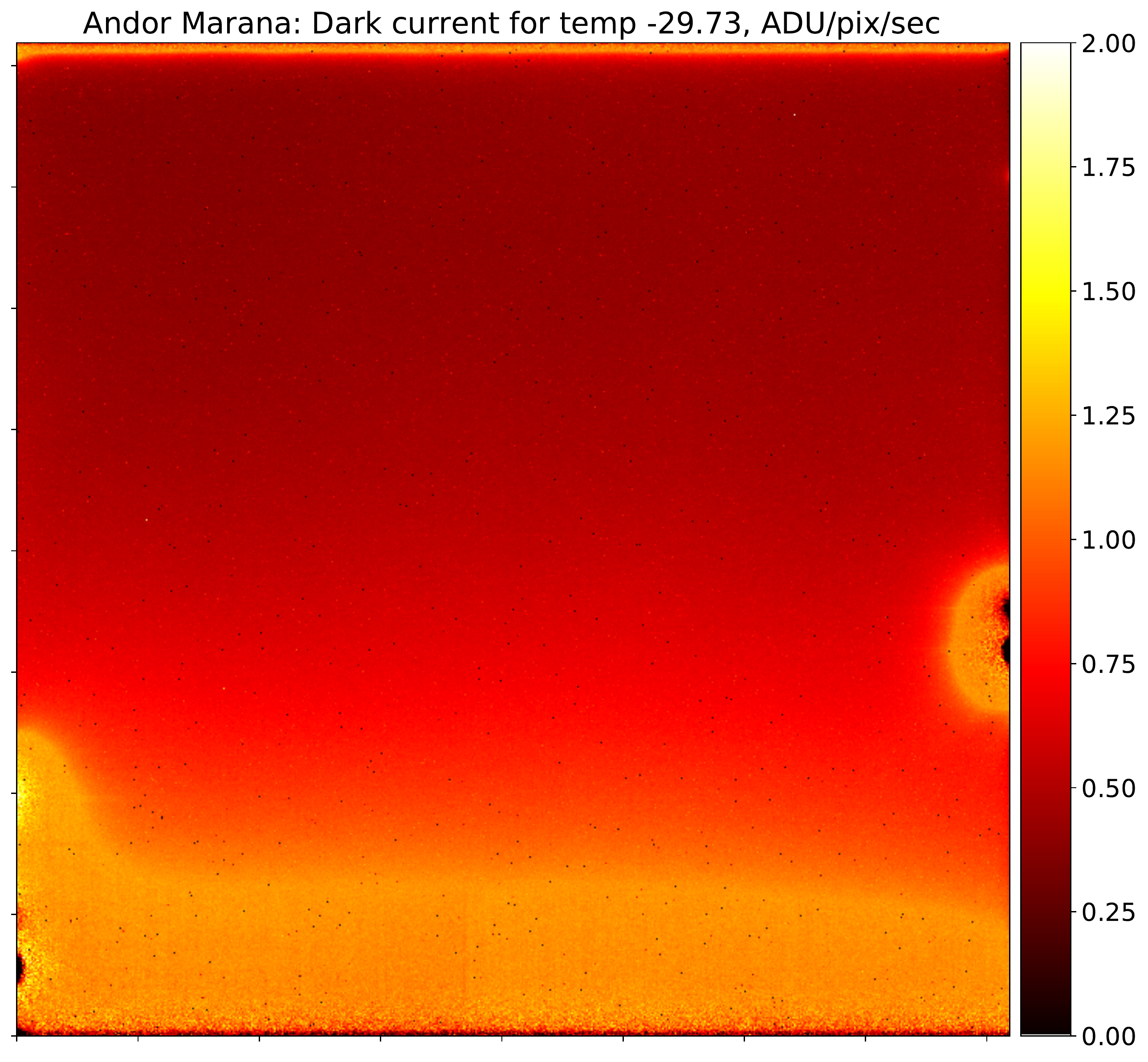}}
    \resizebox*{0.31\columnwidth}{!}{\includegraphics[angle=0]{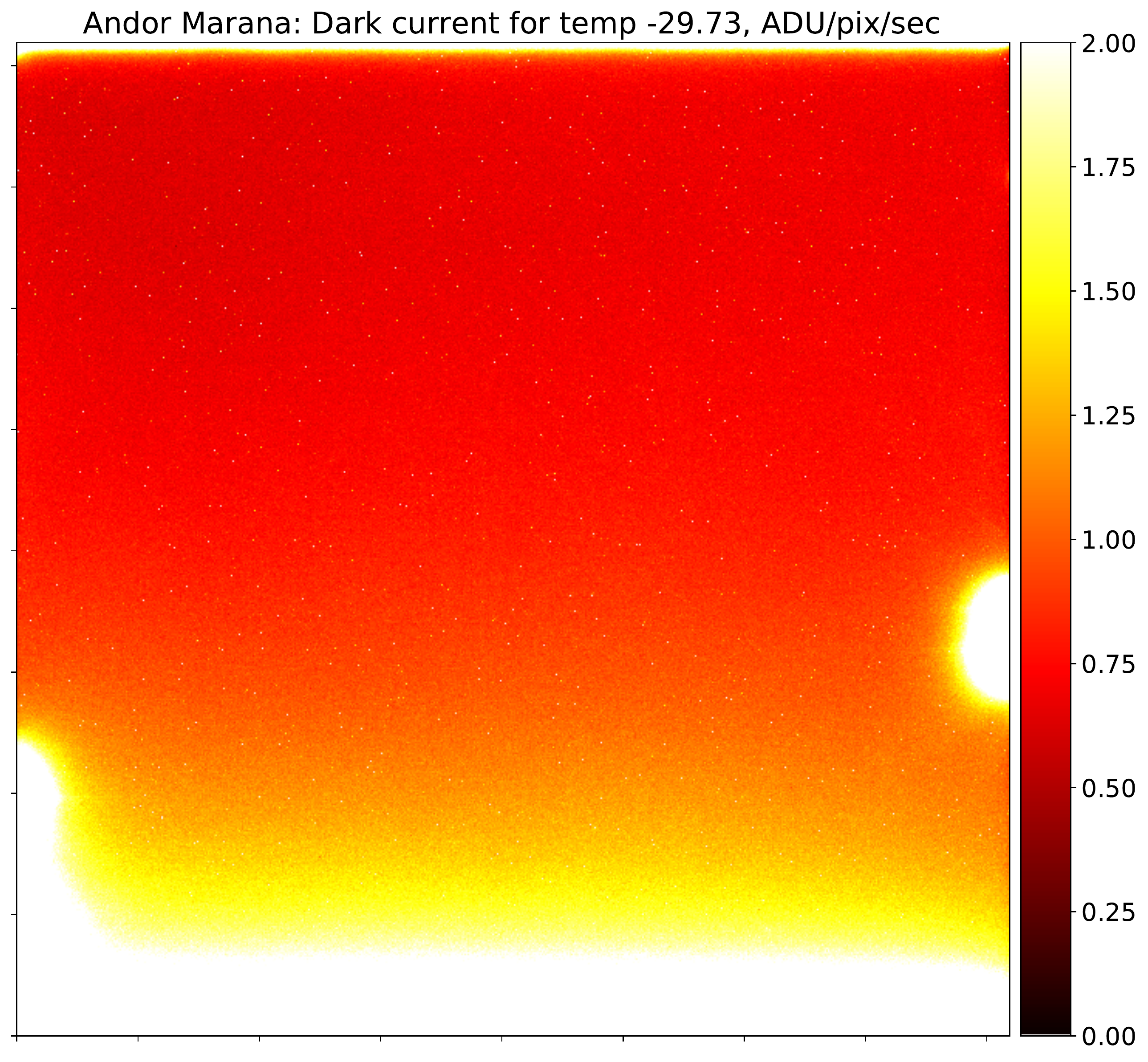}}
  }
  \centerline{
    \resizebox*{0.5\columnwidth}{!}{\includegraphics[angle=0]{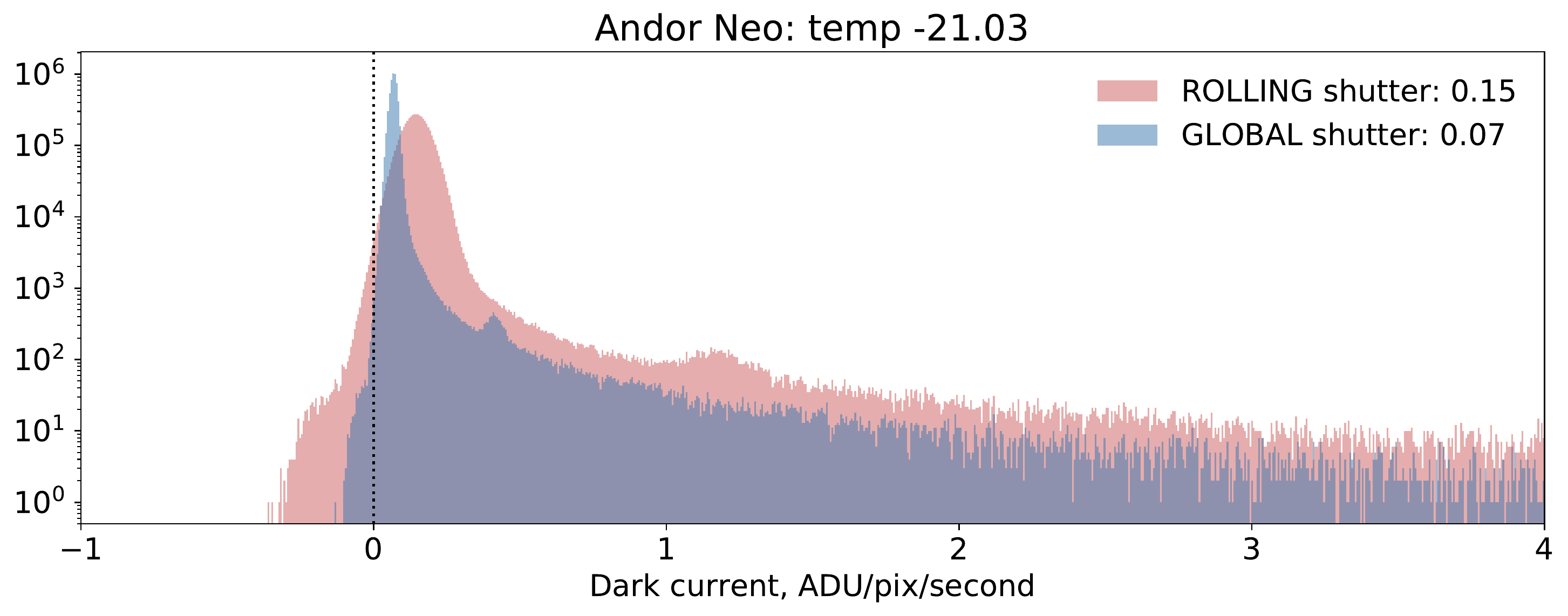}}
    \resizebox*{0.5\columnwidth}{!}{\includegraphics[angle=0]{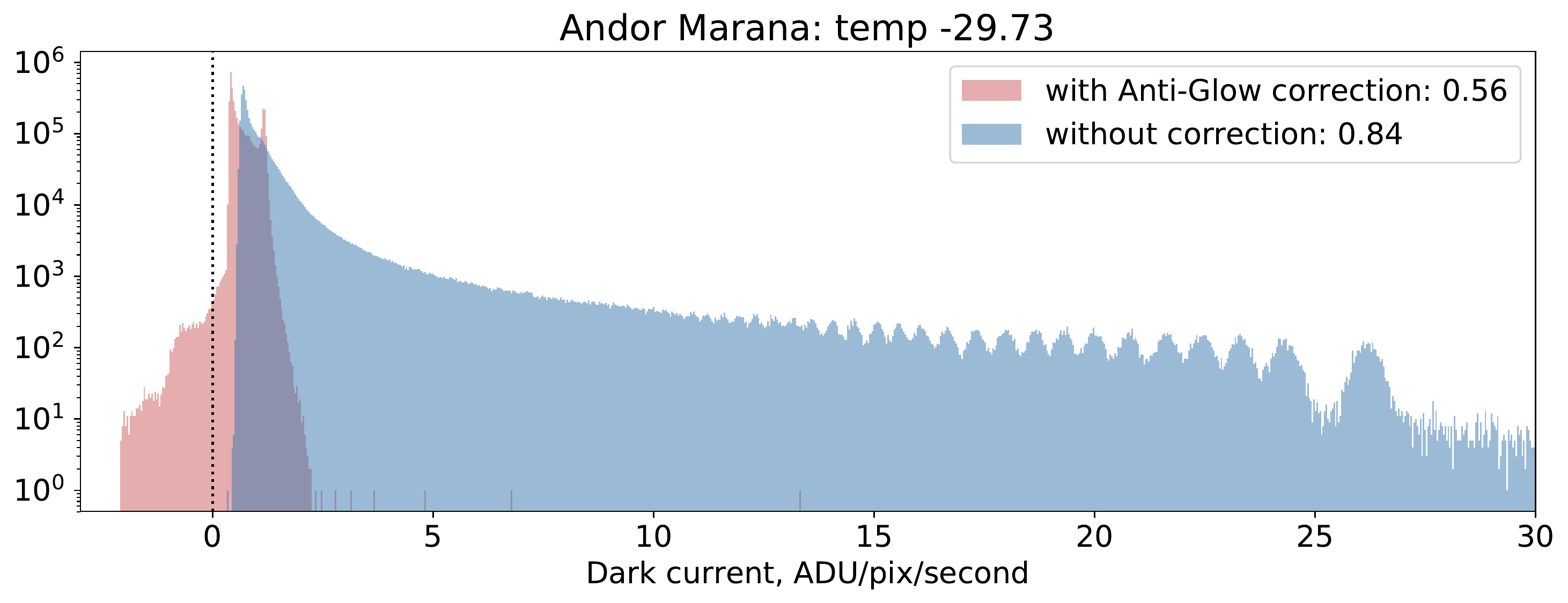}}
  }
  \caption{Maps of a dark current for Andor Neo with global shutter (upper left panel), for Andor Marana with default camera regime with Anti-Glow correction (upper middle panel) and with correction disabled (upper right panel).
    The histograms of dark current for different shutter modes of Andor Neo (lower left panel) and for Andor Marana with and without Anti-Glow correction (lower right panel) The numbers in the legends of lower panels show median values of dark current for corresponding regimes.
    \label{fig_darks}}
\end{figure}

\begin{figure}[t]
  \centerline{
    \resizebox*{0.5\columnwidth}{!}{\includegraphics[angle=0]{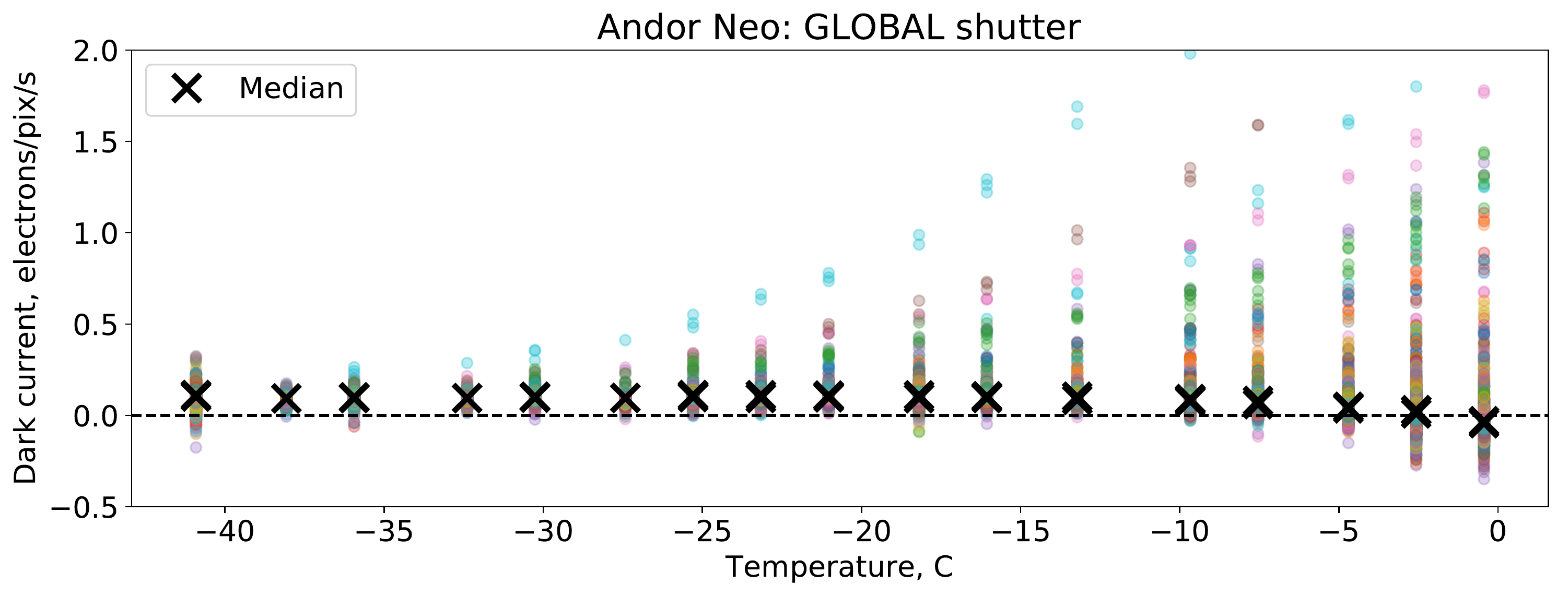}}
    \resizebox*{0.5\columnwidth}{!}{\includegraphics[angle=0]{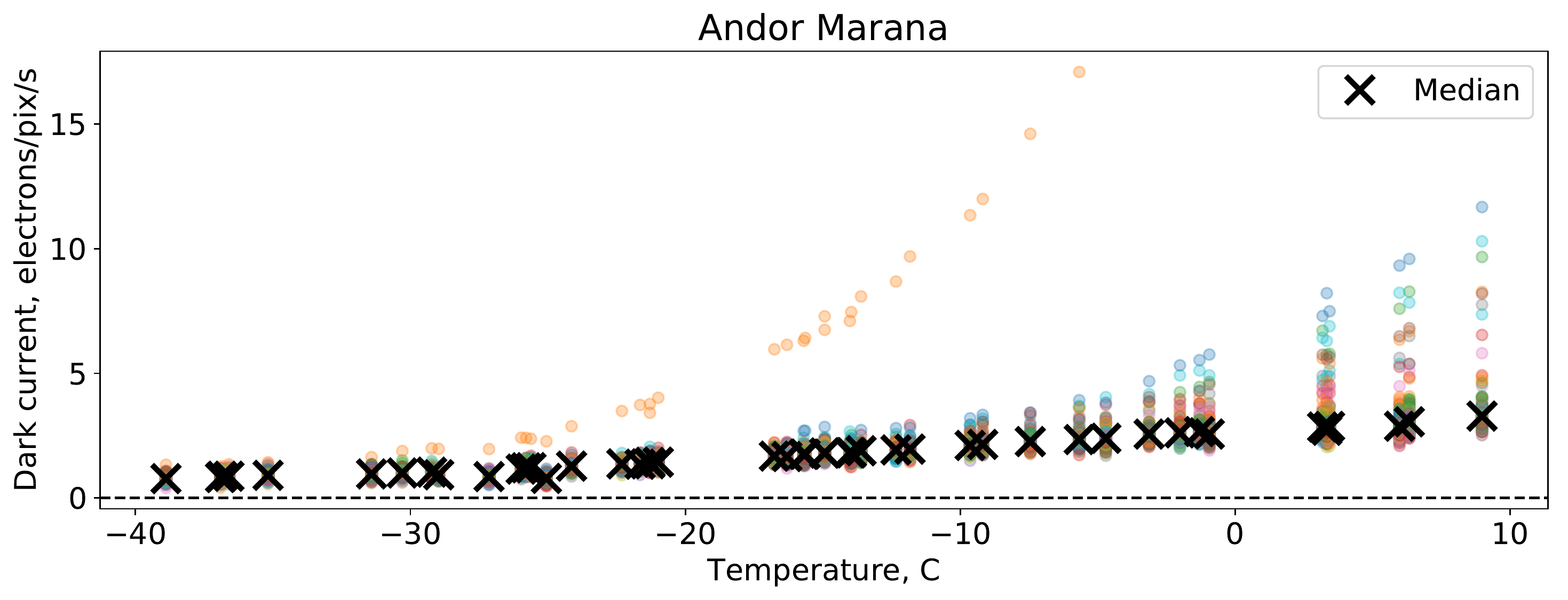}}
  }
  \centerline{
    \resizebox*{0.5\columnwidth}{!}{\includegraphics[angle=0]{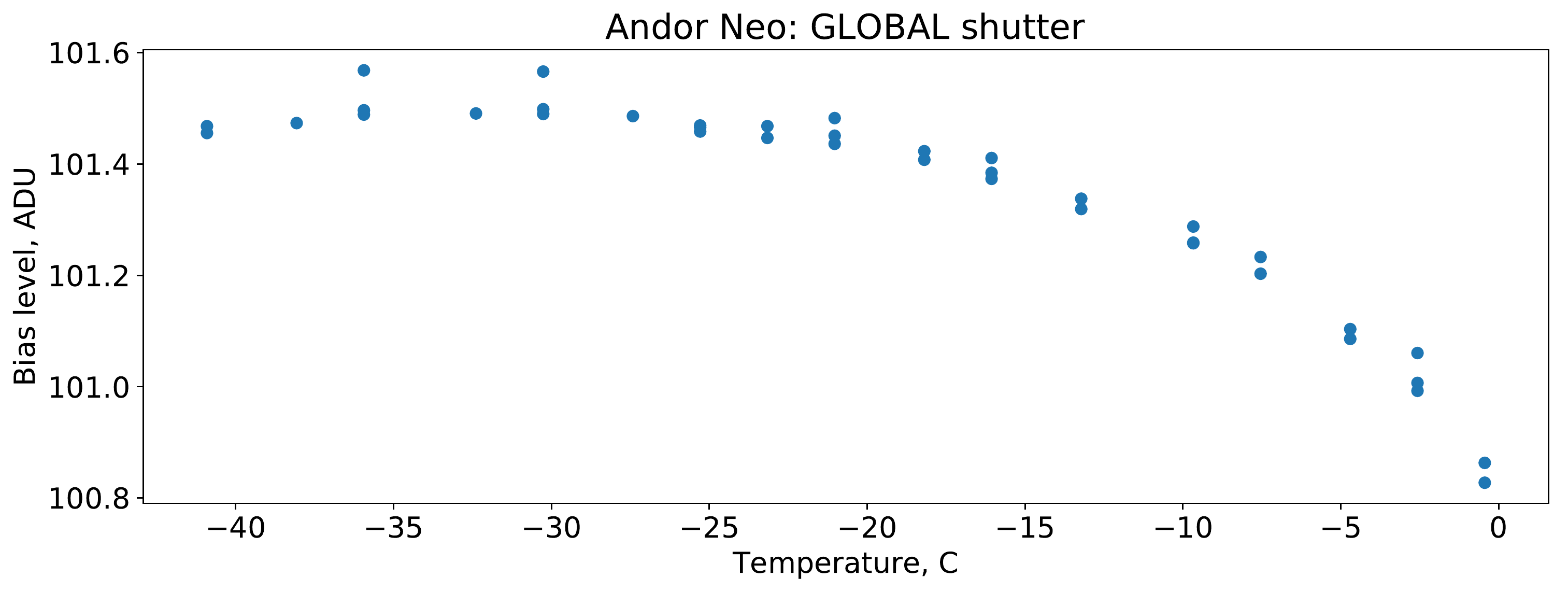}}
    \resizebox*{0.5\columnwidth}{!}{\includegraphics[angle=0]{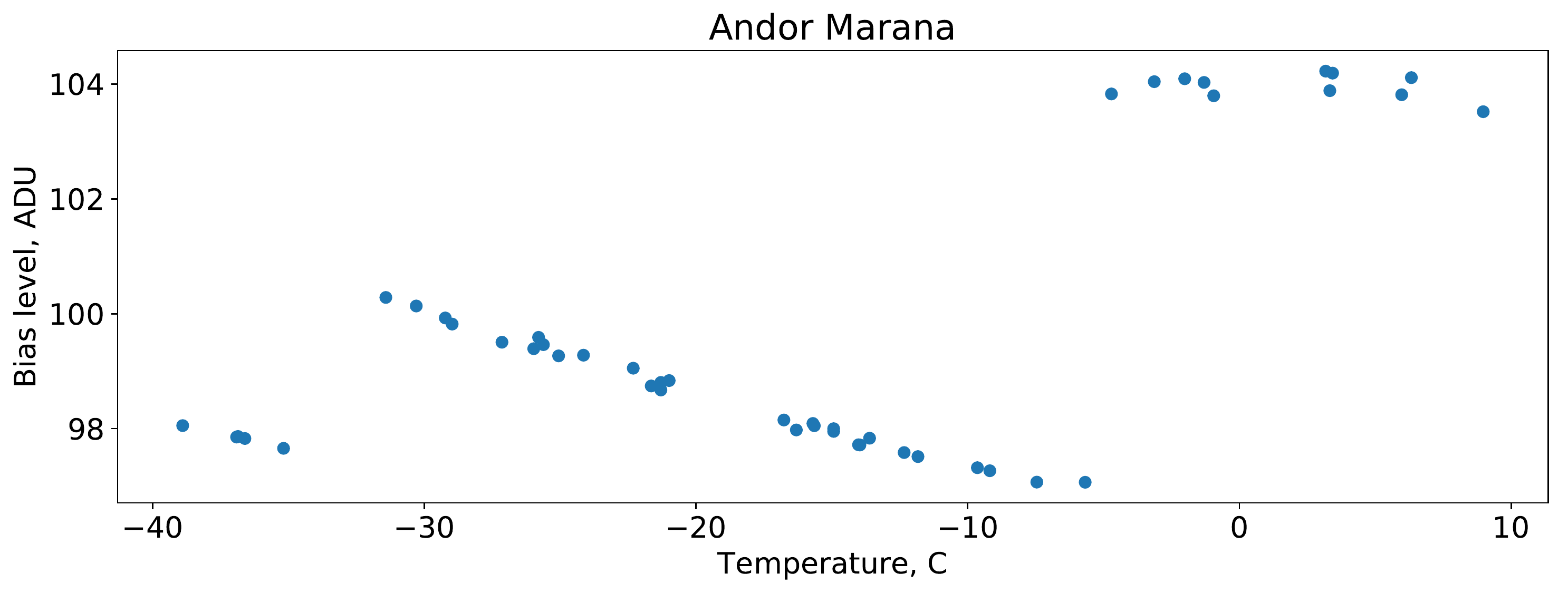}}
  }
  \caption{Temperature dependence of a dark current (upper panels) and bias level (lower panels) estimated for Andor Neo (left) and Andor Marana (right) cameras. A set of representative pixels are randomly chosen in the middle of the frame, far from edge hot spots of Marana. Marana camera has Anti-Glow correction disabled for this test. Bias level corresponds to the median value across set of pixels, while dark current values for every pixel are shown separately, with median overplotted with black crosses.
    \label{fig_darks_temp}}
\end{figure}


Dark current in CMOS imaging sensor is a process of spurious electrons generation in the photodiode in the absence of incoming light, with many possible sources behind it, \textit{e.g.} thermal generation, ``diffusion'' current, etc. To characterize it, we studied series of ``dark'' 
frames acquired with varying exposures. Then the mean value of every pixel was regressed versus exposure time in order to determine both bias level and dark current on a per-pixel basis. The maps of these values are shown in Figure~\ref{fig_darks}. While for Neo the dark current is mostly flat and has a median of about 0.1 e$^-$/pix/s, for Marana it shows quite significant edge glow towards both top and bottom edges of the sensor, as well as an extended hot spots along the vertical edges, with median value of 0.8 e$^-$/pix/s. Interestingly, for 6429 pixels (0.15\% of all) formally measured dark current is negative, as the dark pixel value linearly drops with increased exposure until reaches some fixed level. We attribute it to the occasional over-compensation of a dark current due to  application of ``Anti-Glow technology'' algorithms during on-board frame processing. Turning off this correction (using an undocumented \texttt{GlowCompensation} Software Development Kit (SDK) option) removed such negative values from the dark current map and significantly increased the amplitude of edge glow spots, leading to a long tail in the histogram of its values (see lower right panel of Figure~\ref{fig_darks} for a comparison of histograms with and without glow correction) and increasing the dark current median value up to 1.2 e$^-$/pix/s. At the same time, the amount of hot pixels with dark current 10 times greater than median value is reaching 2\% for Marana with Anti-Glow correction disabled, while it is just 0.1\% for Neo.


Temperature dependence of the dark current, along with effective bias level, are shown in Figure~\ref{fig_darks_temp} for both cameras. While for Marana the dark current predictably grows with temperature for all pixels, for Neo it stays nearly constant above -10$^\circ$, and then drops towards negative values for the majority of pixels as temperature rises towards 0$^\circ$. On the other hand, for Marana the bias level displays distinctive jumps around -5$^\circ$ and -32$^\circ$ suggesting that some compensatory mechanism is still in effect there, despite Anti-Glow correction being turned off. Probably, similar mechanism also explains unexpected behaviour of dark current for Neo camera, if it somehow depends on the exposure time. Thus we have to avoid drawing any physical conclusions about the sources of dark current, like activation energy etc, based solely on our experimental data.


\section{Pixel noise}

\label{sec_pixelstability}


\begin{figure}[t]
  \centerline{
    \resizebox*{0.5\columnwidth}{!}{\includegraphics[angle=0]{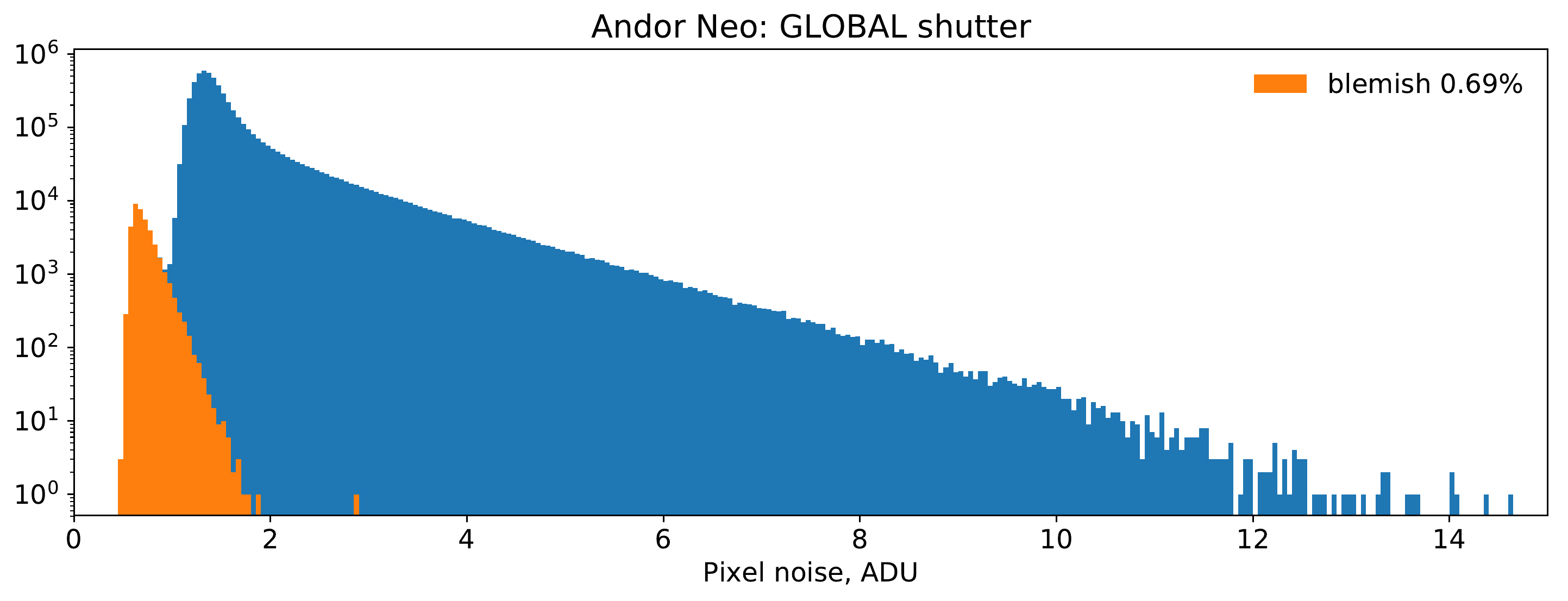}}
    \resizebox*{0.5\columnwidth}{!}{\includegraphics[angle=0]{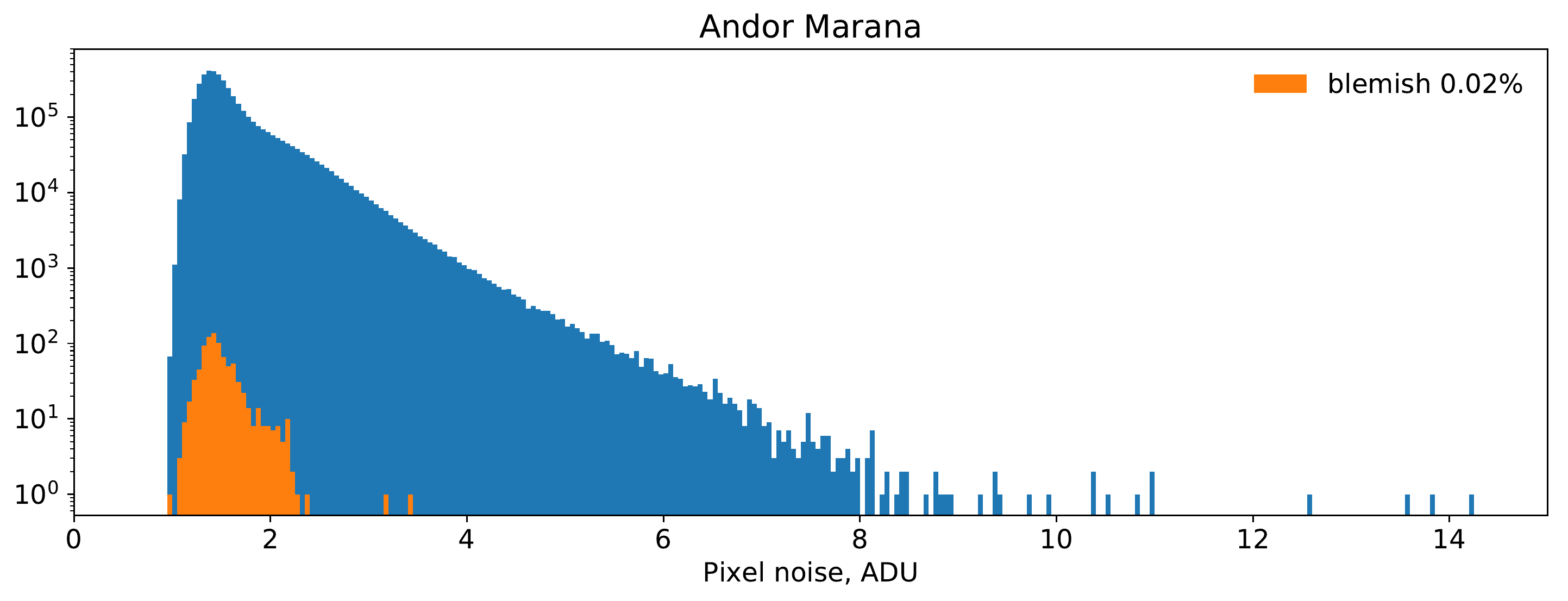}}
  }
  \caption{Histogram of a pixel noise RMS on a sequence of dark frames. Blemished pixels are masked and being replaced on the fly by onboard FPGA logic with the mean value of a pixels surrounding them, and thus effectively have smaller noise amplitude.
    \label{fig_noise_dark}}
\end{figure}

\begin{figure}[t]
  \centerline{
    \resizebox*{0.5\columnwidth}{!}{\includegraphics[angle=0]{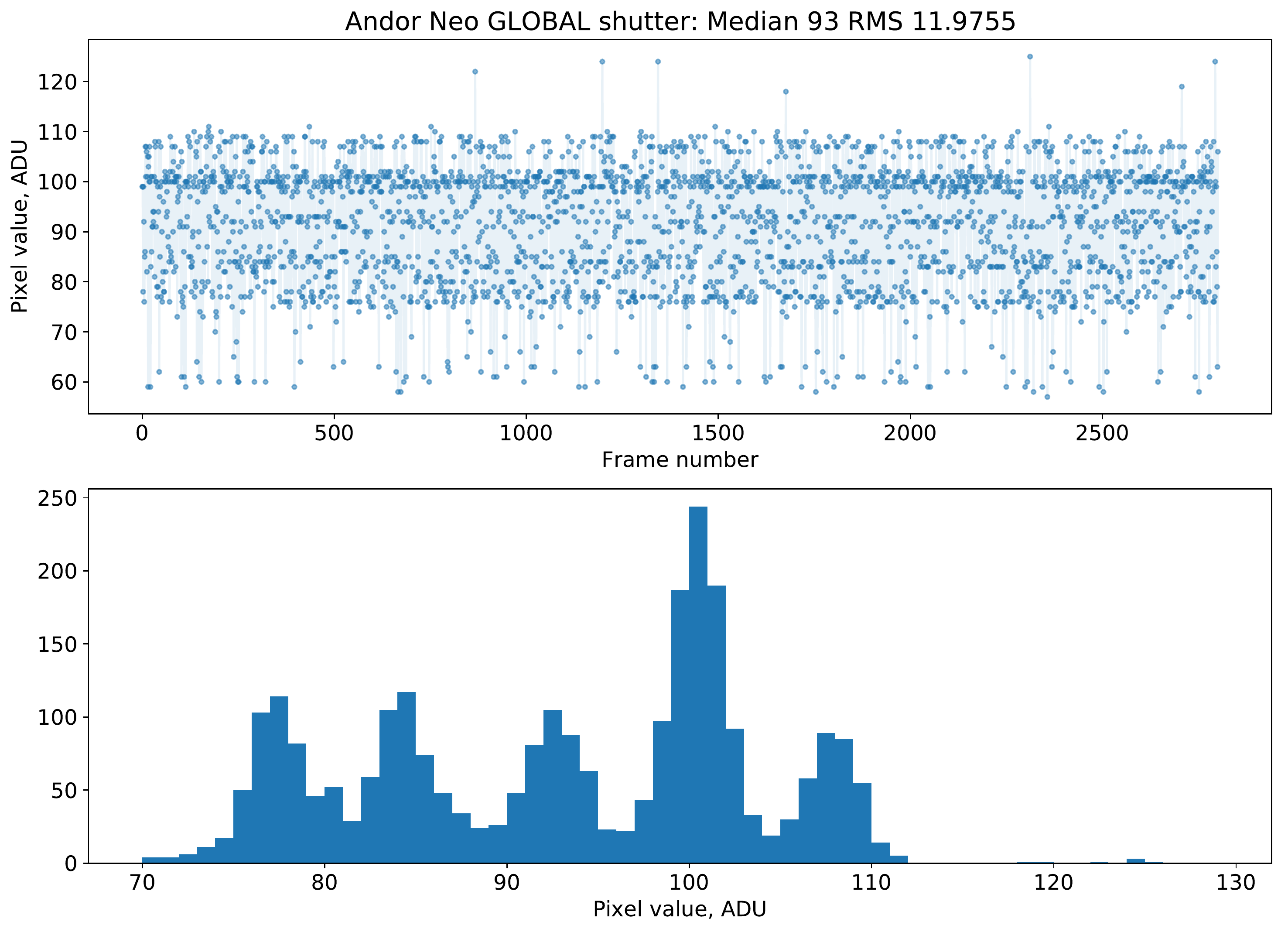}}
    \resizebox*{0.5\columnwidth}{!}{\includegraphics[angle=0]{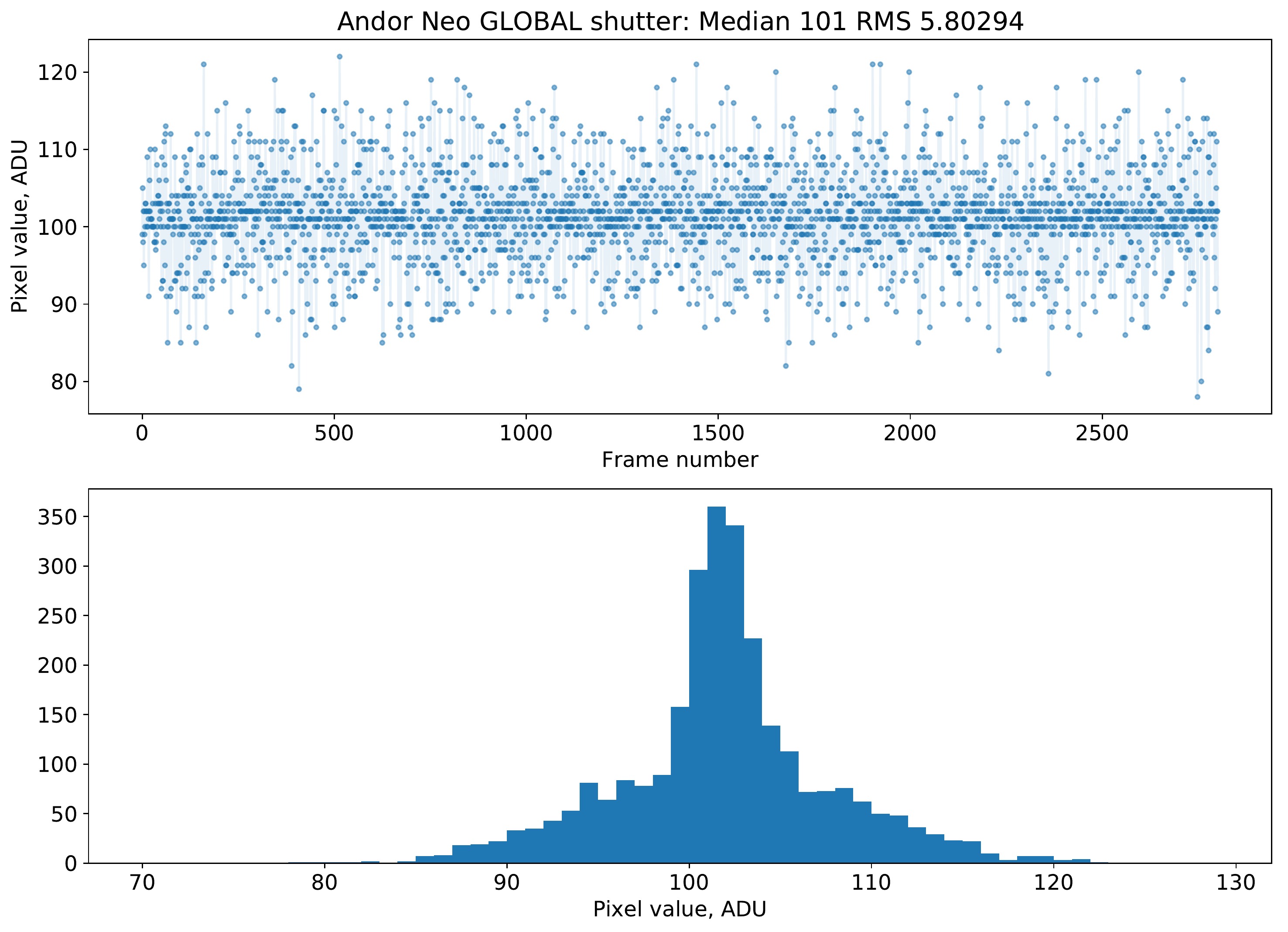}}
  }
  \caption{Examples of noisy pixels that displays distinctive Random Telegraph Signal (RTS) switching between several bias states on a dark frames due to charge traps in source follower of a pixel. Upper panels -- temporal sequence of pixel values. Lower panels -- histogram of these values.
    \label{fig_noise_rts}}
\end{figure}

\begin{figure}[t]
  \centerline{
    \resizebox*{0.2\columnwidth}{!}{\includegraphics[angle=0]{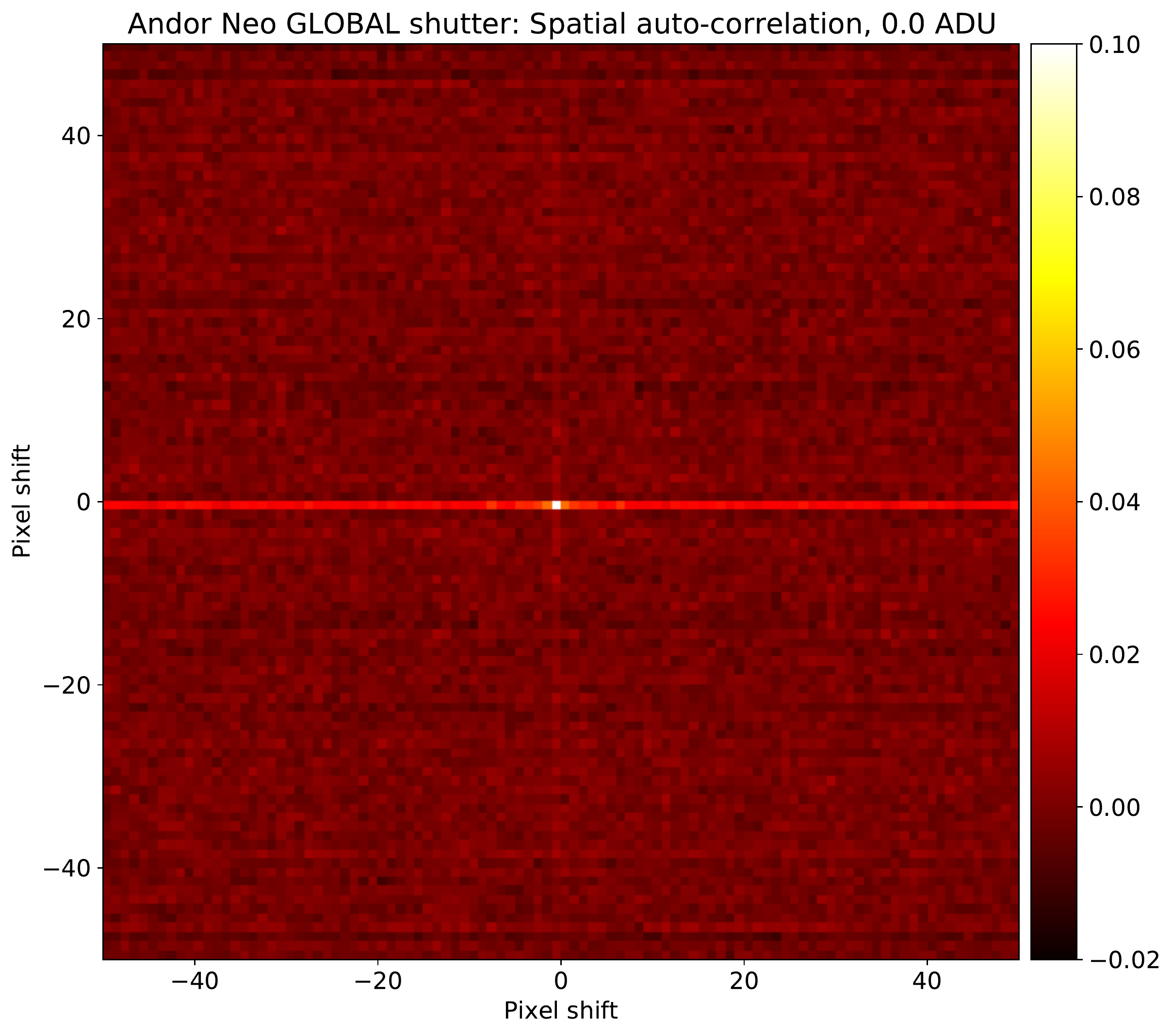}}
    \resizebox*{0.2\columnwidth}{!}{\includegraphics[angle=0]{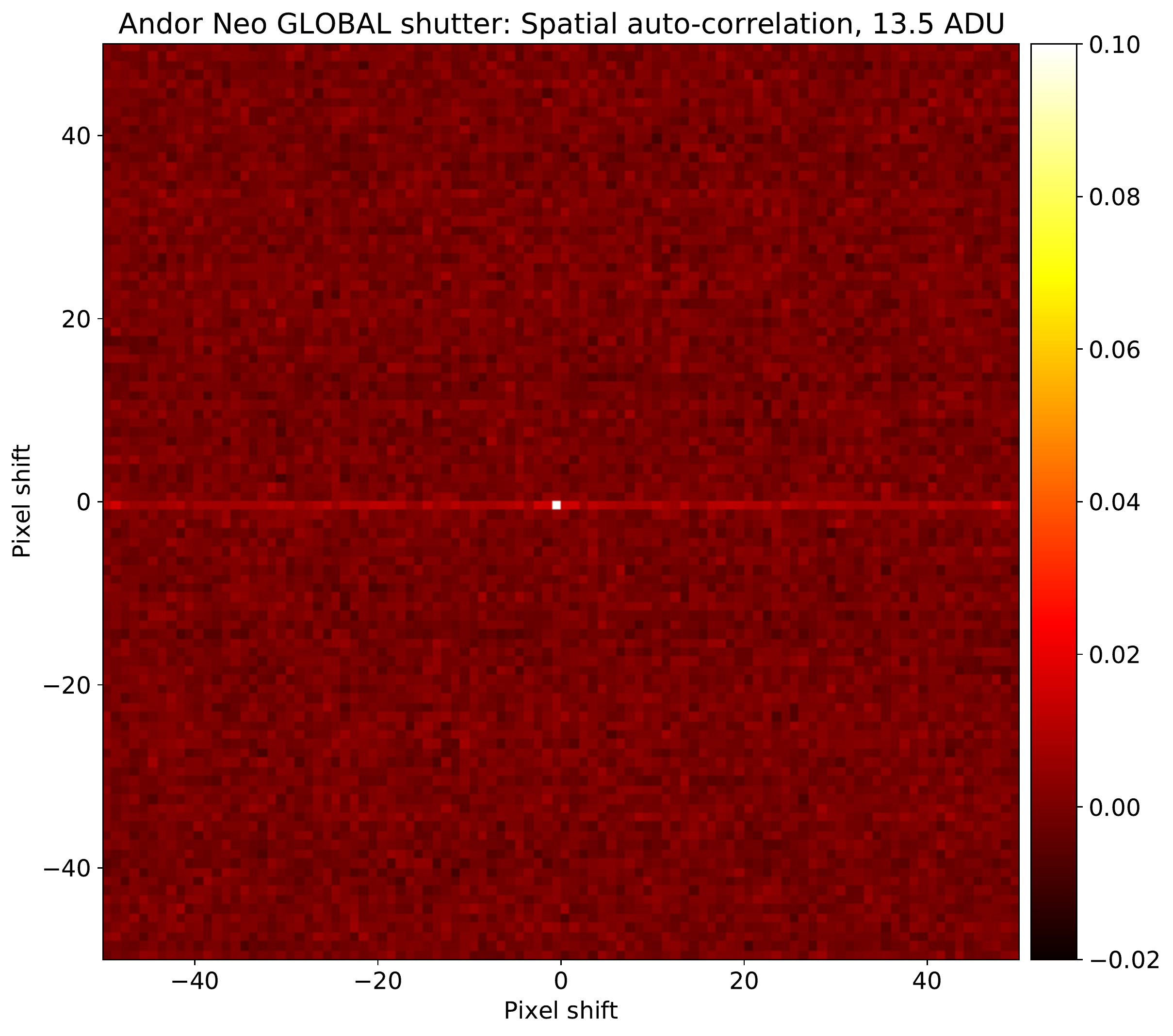}}
    \resizebox*{0.2\columnwidth}{!}{\includegraphics[angle=0]{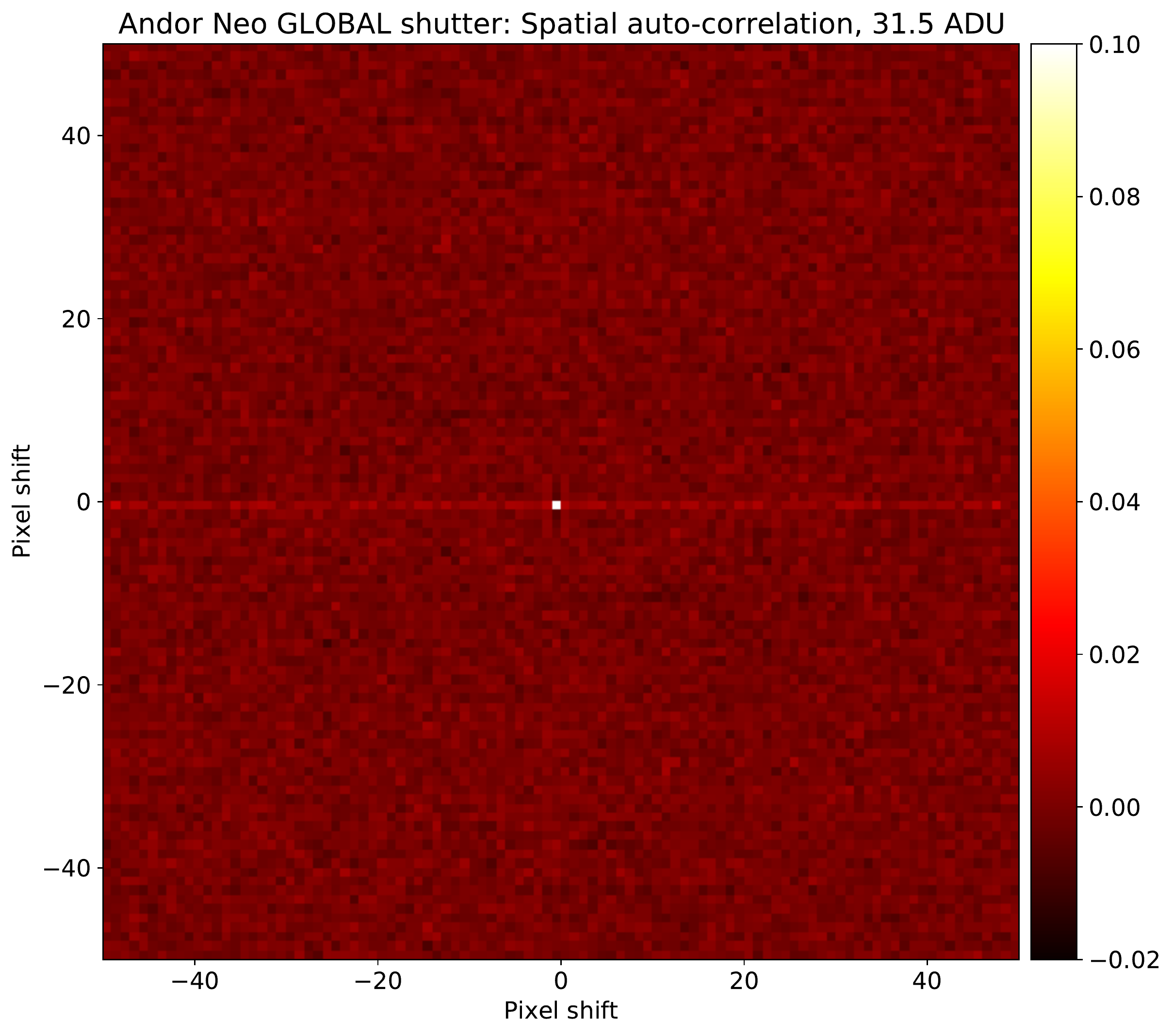}}
    \resizebox*{0.2\columnwidth}{!}{\includegraphics[angle=0]{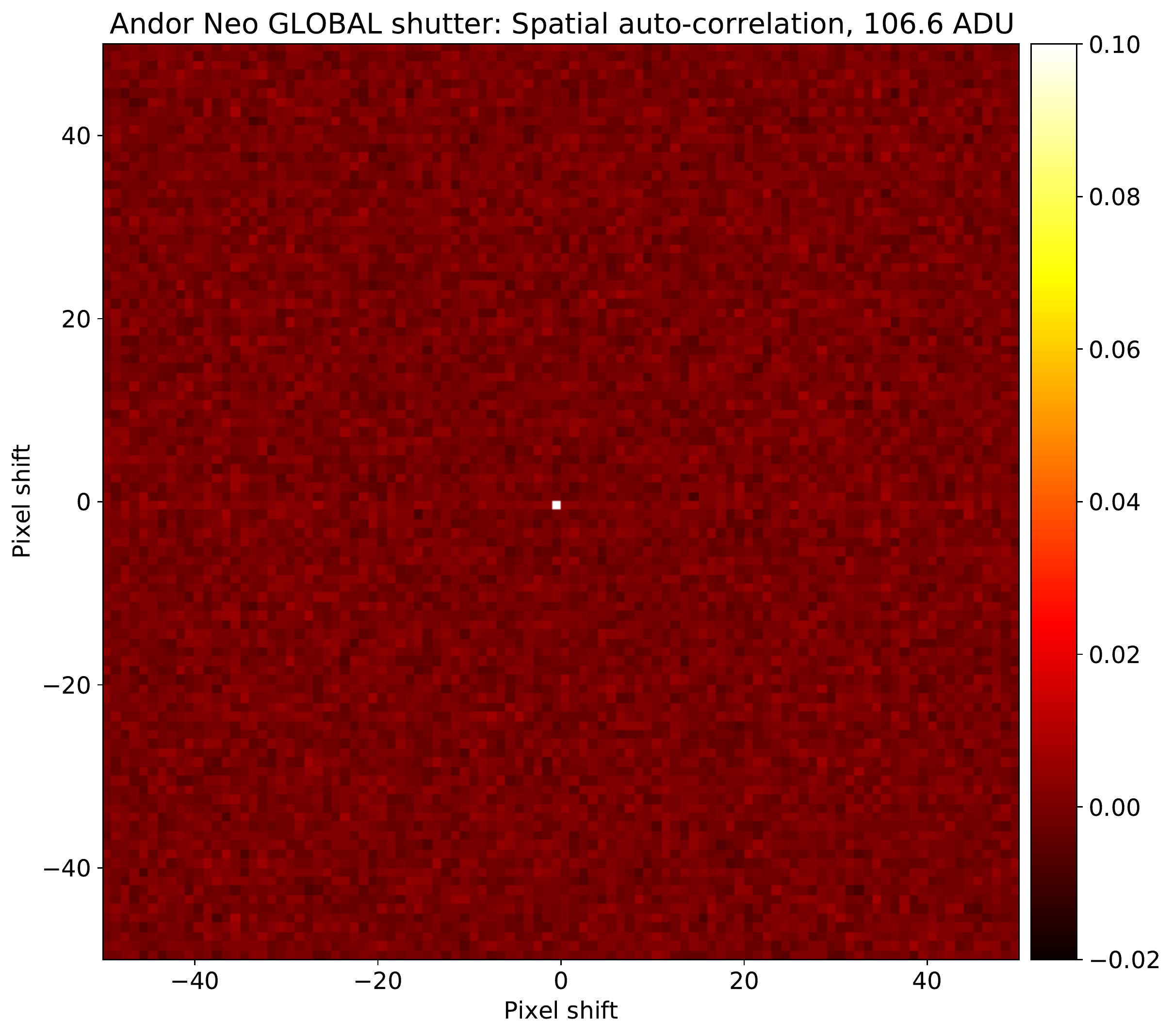}}
    \resizebox*{0.2\columnwidth}{!}{\includegraphics[angle=0]{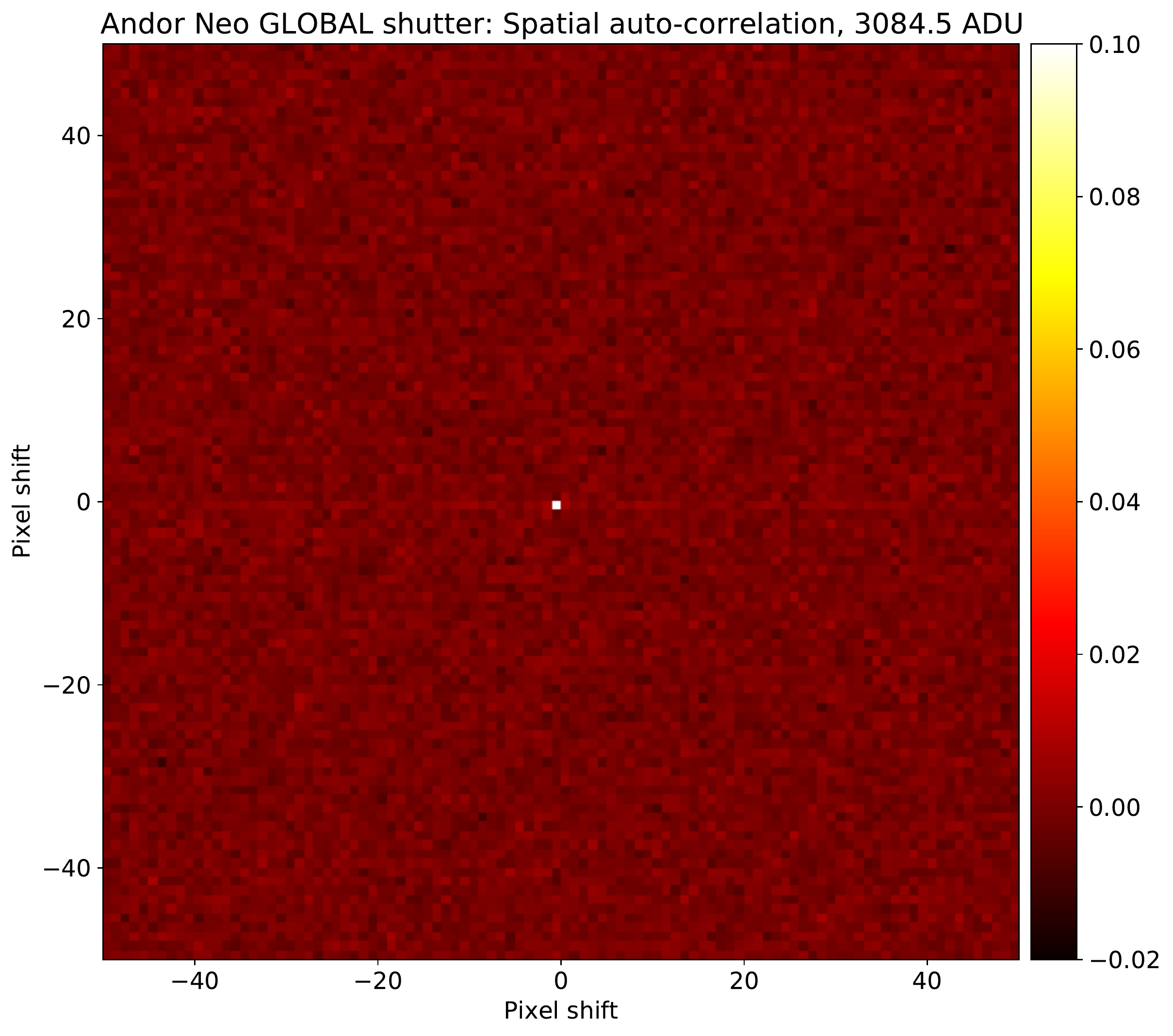}}
  }
  \centerline{
    \resizebox*{0.2\columnwidth}{!}{\includegraphics[angle=0]{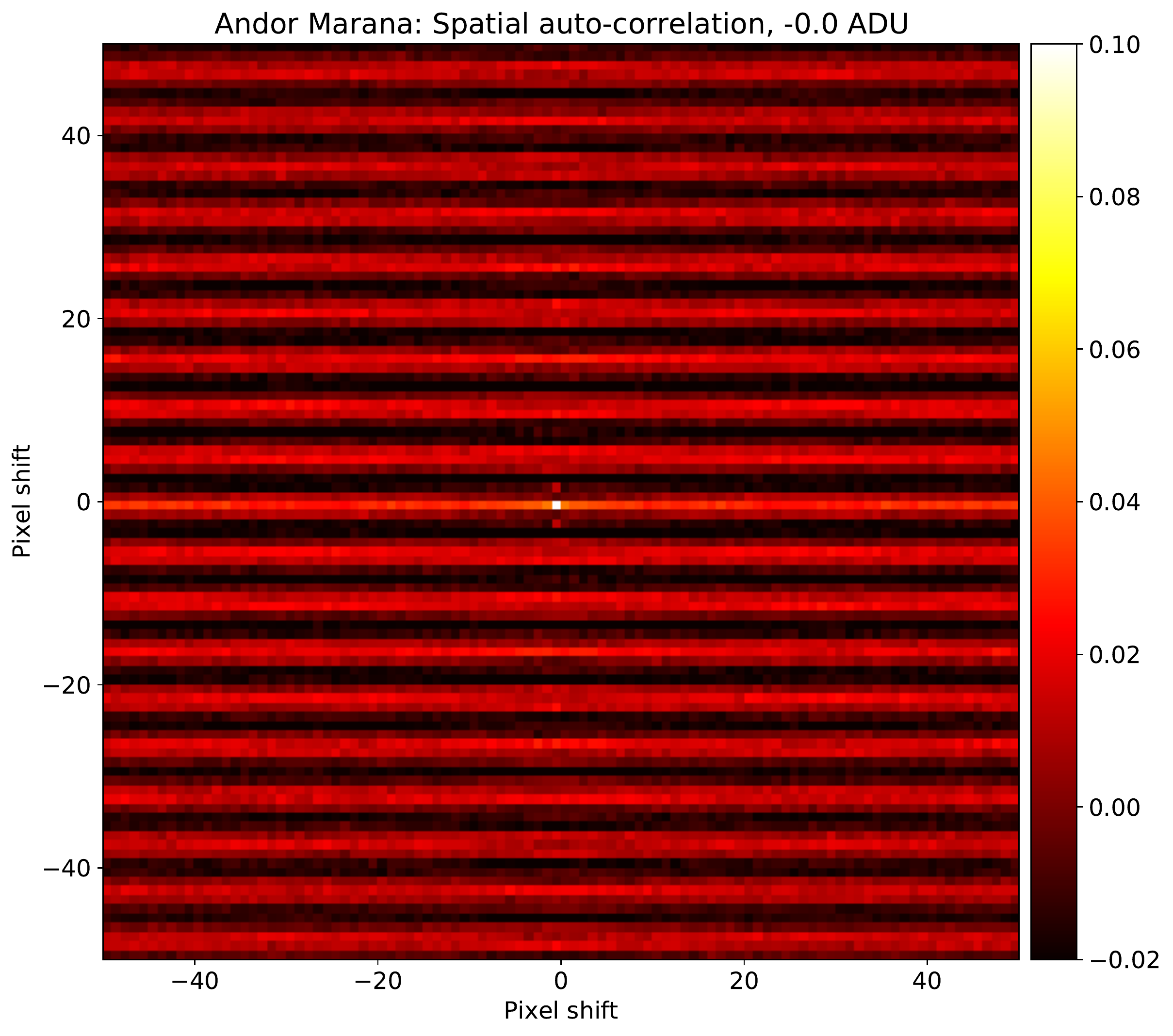}}
    \resizebox*{0.2\columnwidth}{!}{\includegraphics[angle=0]{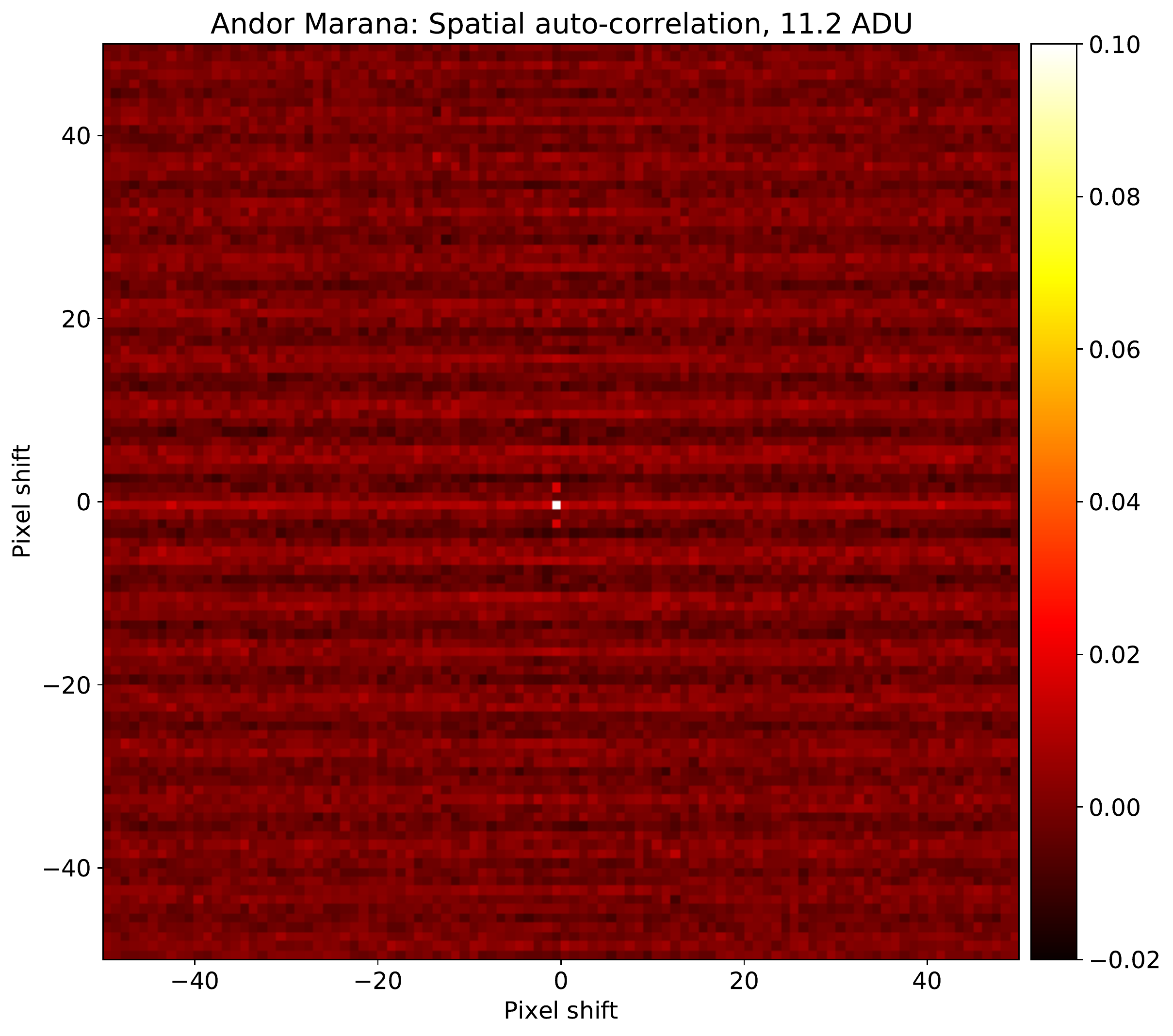}}
    \resizebox*{0.2\columnwidth}{!}{\includegraphics[angle=0]{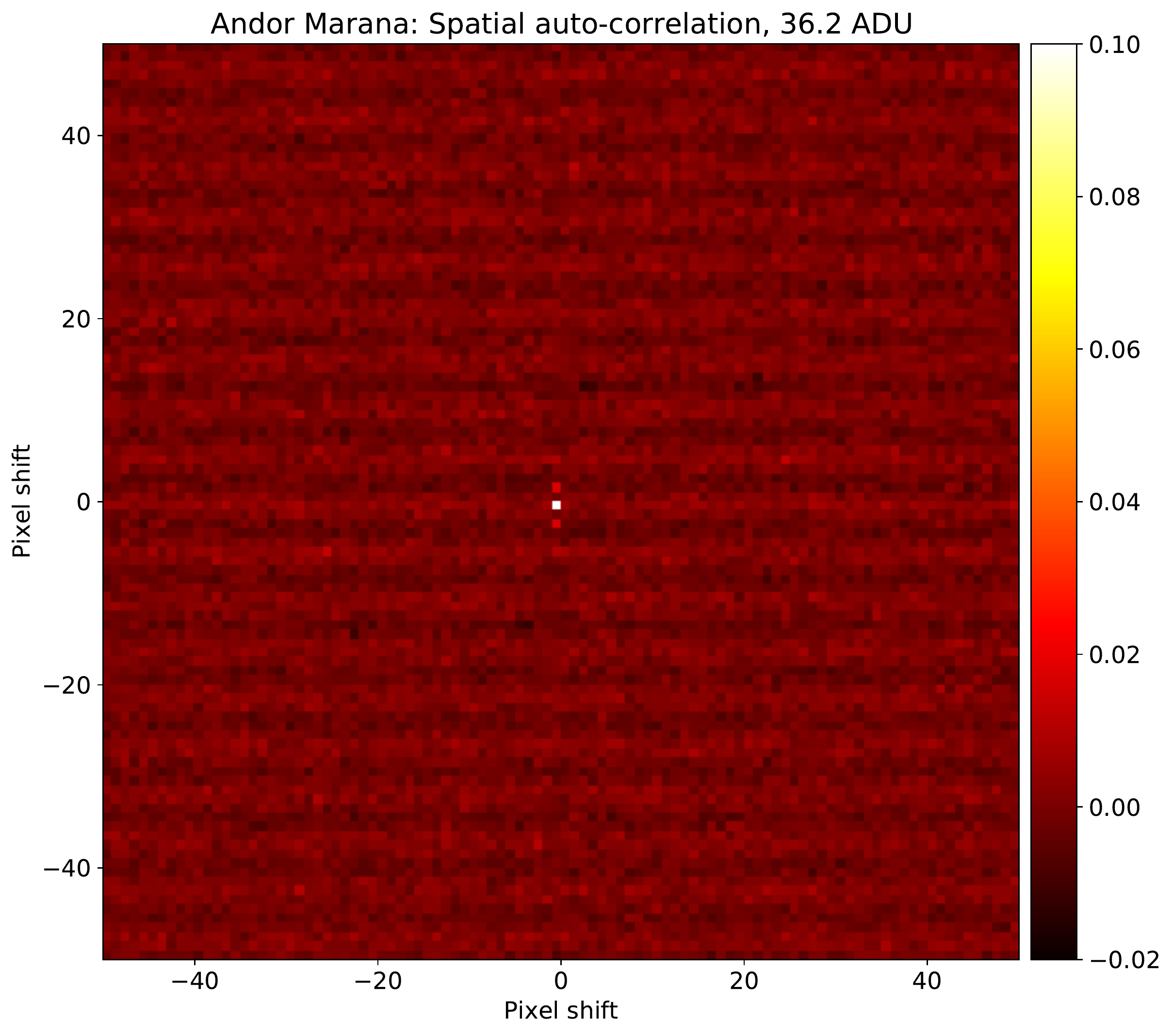}}
    \resizebox*{0.2\columnwidth}{!}{\includegraphics[angle=0]{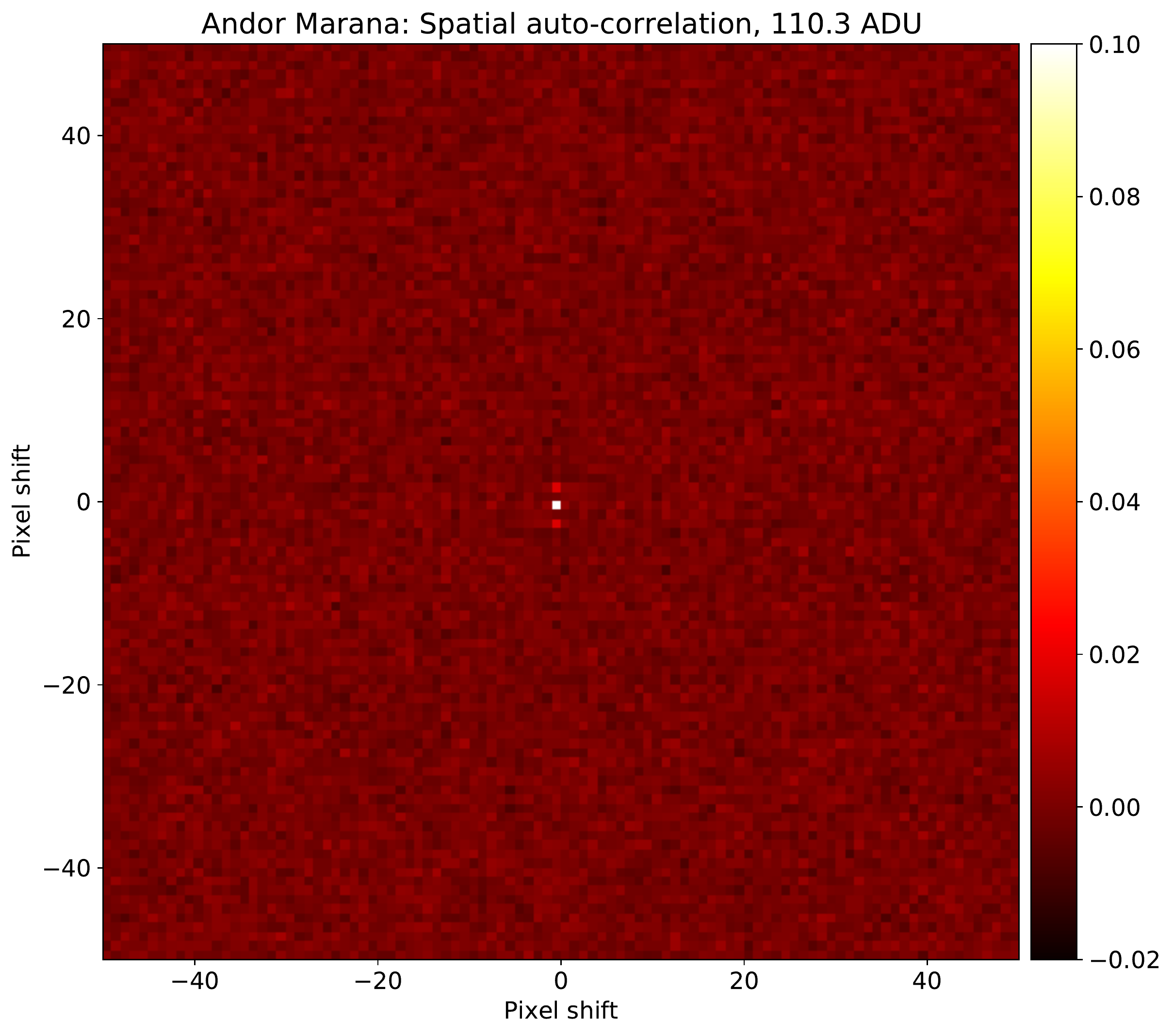}}
    \resizebox*{0.2\columnwidth}{!}{\includegraphics[angle=0]{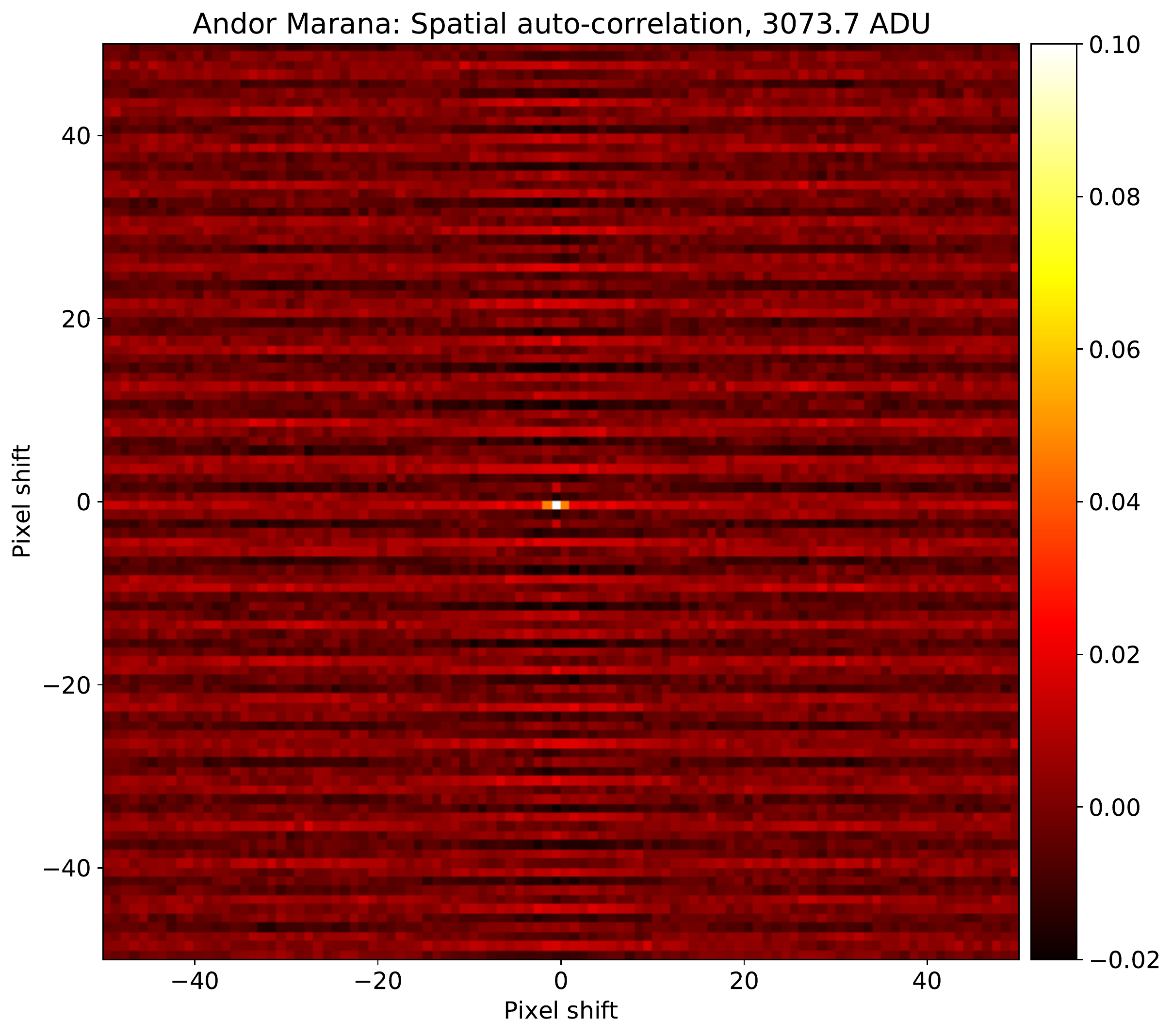}}
  }
  \caption{Spatial auto-correlation of a pixel values on individual frame after subtraction of mean value over a sequence (i.e. fixed pattern noise component) for different intensities of incoming light on Andor Neo (upper row) and Andor Marana (lower row) cameras. All frames are scaled to the same intensity mapping covering the range from -0.1 to 0.1. On Andor Neo, the correlation along the lines is apparent and is most probably related to limited accuracy of onboard subtraction of overscans (overscan pixels are physically present on CIS2521 chip, but not accessible for an end user through Andor SDK). The influence of this correlated noise component quickly fades with the increase of intensity of incoming light and is completely invisible above $\sim$100 ADU.
    Andor Marana shows more complex distinctive striped pattern corresponding also to some (anti-)correlation of consecutive rows, which we can't readily interpret. It also fades away above $\sim$100 ADU, but re-appears above the level of transition to high gain amplifiers.
    \label{fig_spatial_corr}}
\end{figure}


In order to study the noise behaviour on a per-pixel basis, we analyzed the behaviour of every pixel values over long consecutive sequences of frames, both dark and illuminated. The first point to note about the noise properties on Andor sCMOS cameras is an interpolative masking of a small subset of ``blemished'' pixels having either too high dark current or excess read-out noise. An on-board ``blemish correction'' \cite{neo_blemish} replaces the values of these pixels with an average of 8 surrounding ones on the fly, effectively leading to formation of a sub-set of pixels with noise level on average $\sqrt{8}$ times smaller than the rest. Direct search for such pixels may be performed by analyzing a series of illuminated images and locating the pixels with values always equal to arithmetic mean of its surroundings. The analysis of acquired Marana data allowed us to identify 0.02\% of pixels as a blemish masked, mostly concentrated towards the edges of the chip. The same  value for Andor Neo is around 0.66\%, and the pixels are randomly scattered across the whole frame.

The noise on individual frames from Andor Neo after subtraction of mean pixel level is slightly correlated along the rows (see Figure~\ref{fig_spatial_corr}) which we attribute to the effect of onboard compensation of row to row baseline variations using overscan pixels. On the other hand, the noise on Andor Marana frames show distinct anti-correlation between adjacent rows which we can't readily interpret. Both of these effects lead to a slightly ``striped'' images at low light intensities, and gradually disappear as the intensity grows. However, the striped pattern is again apparent on Marana frames where the signal is processed using high-gain amplifiers (i.e. having intensity above $\sim$1500 ADU).

\begin{figure}[t]
  \centerline{
    \resizebox*{0.5\columnwidth}{!}{\includegraphics[angle=0]{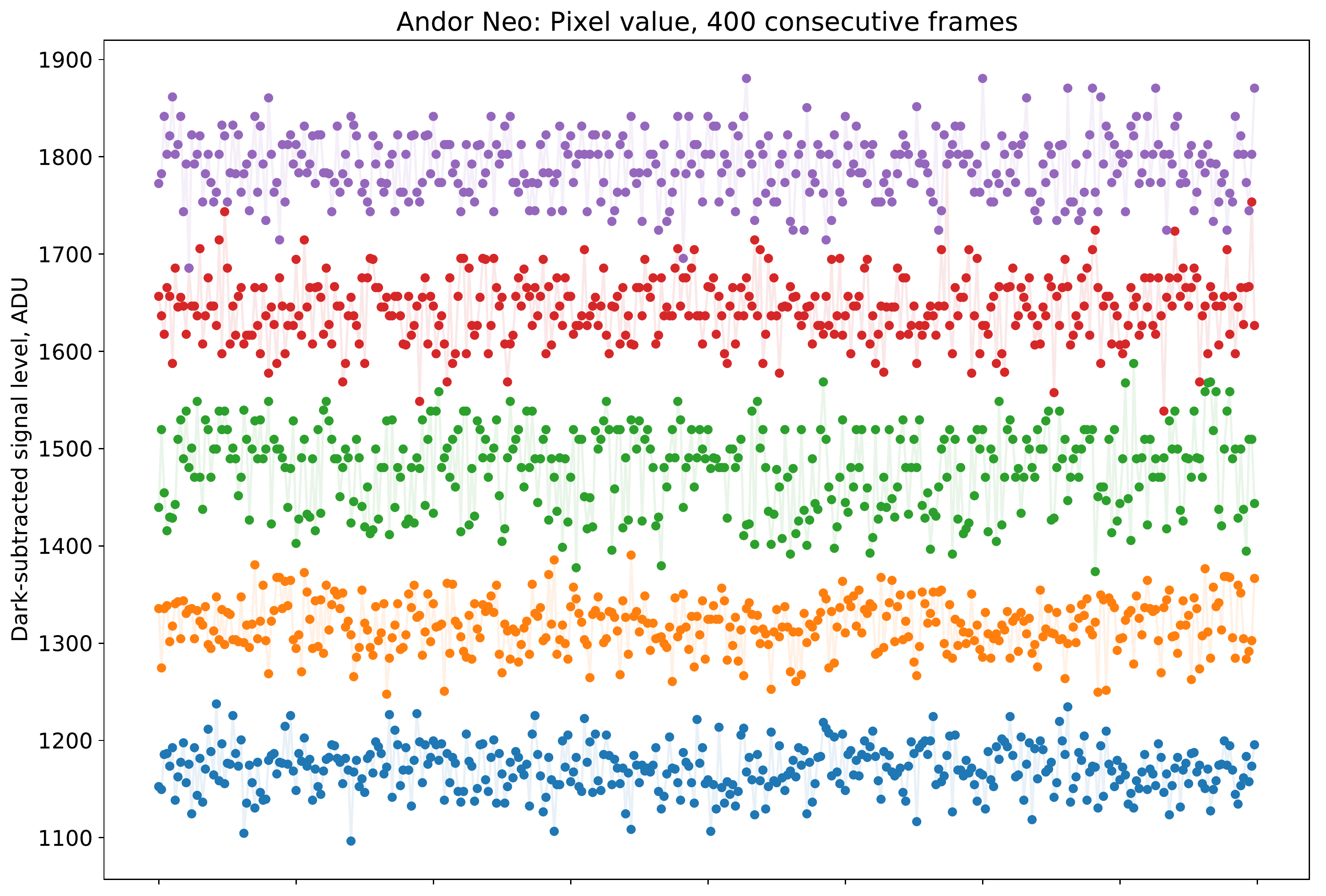}}
    \resizebox*{0.465\columnwidth}{!}{
      \begin{tabular}[b]{l}
        \resizebox*{0.5\columnwidth}{!}{\includegraphics[angle=0]{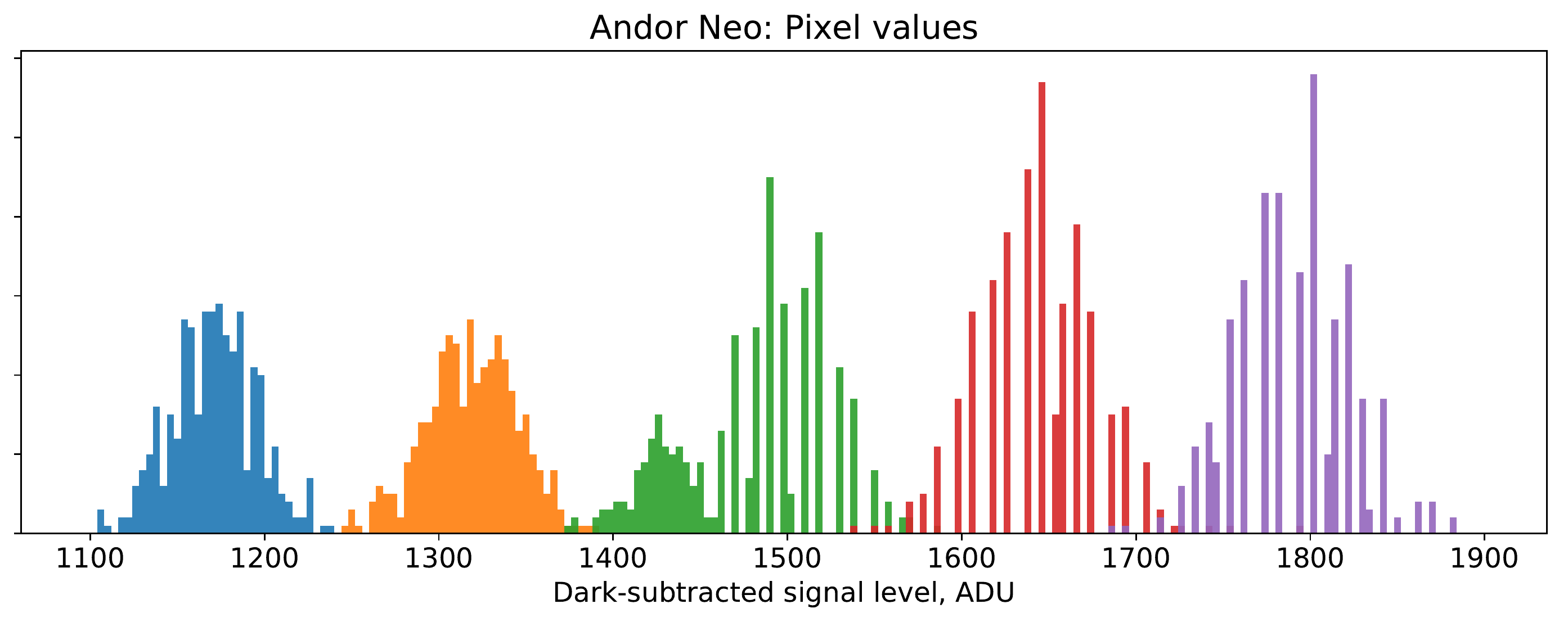}}\\
        \resizebox*{0.5\columnwidth}{!}{\includegraphics[angle=0]{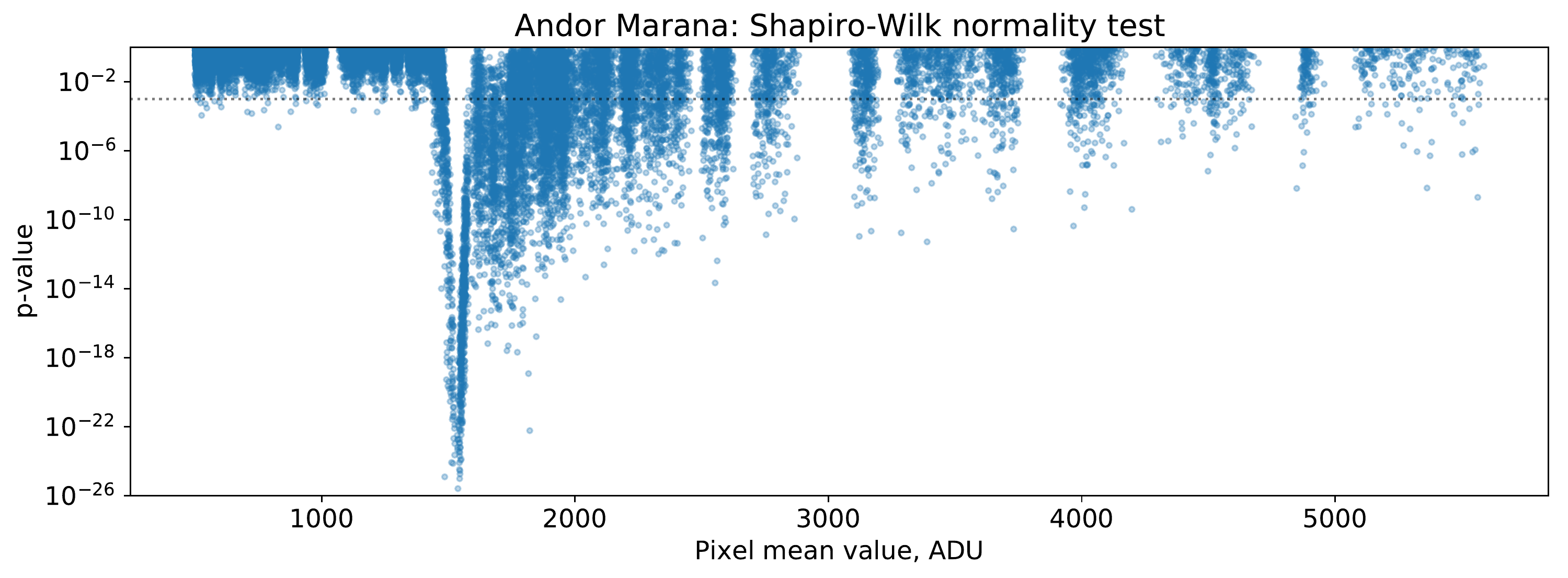}}
      \end{tabular}
    }
  }
  \caption{Behaviour of individual pixel values around the intensity region where transition between low gain and high gain amplifiers happen. Left panel -- sequences of values for the same pixel under slightly different intensities of illumination. Upper right panel -- histograms of these values, showing distinct change to a discrete distribution of values for intensities above $\sim$1500 ADU, corresponding to values being upscaled from a lower-resolution readings from high-gain amplifiers. Behaviour of values above amplifier transition on Andor Marana does not show such a distinctive discrete levels. Lower right panel -- p-values of Shapiro-Wilk normality test applied to sequences of pixel values of different intensities on Andor Marana, showing a rapid transition to non-Gaussian behaviour above the transition. Pixel values distribution remain significantly non-Gaussian up to mean intensities of $\sim$3000-4000 ADU.
    \label{fig_noise_transition}}
\end{figure}

The histograms of a noise on dark frames (see Figure~\ref{fig_noise_dark}) show a significant power-law tail towards high RMS values, which consists of a pixels often displaying distinctive Random Telegraph Signal (RTS) features, effectively ``jumping'' between several metastable signal levels due to effect on electron traps inside the pixel circuits (most often source follower) during either first or second readout during the correlated double sampling process \cite{noise_rts}. The example of such behaviour is shown in Figure~\ref{fig_noise_rts}.
Under illumination levels with Poisson noise exceeding the amplitude of these jumps (which is typically around 10 ADU for both Neo and Marana cameras), these pixels behave normally and their histograms does not show any significant deviations from a Gaussian.
Then, on approaching the amplifier transition region (see photon transfer curves for both cameras in Figure~\ref{fig_ptc}), another effect appears, related to the switching between the readings of low gain and high gain amplifiers, effectively looking like a spontaneous jumping of the value -- see Figure~\ref{fig_noise_transition} for an example. Further increase of the intensity leads to only high gain readings being used, and pixel values are again stable.



\subsection{Linearity and pixel response non-uniformity}\label{sec_pts}

\begin{figure}[t]
  \centerline{
    \resizebox*{0.5\columnwidth}{!}{\includegraphics[angle=0]{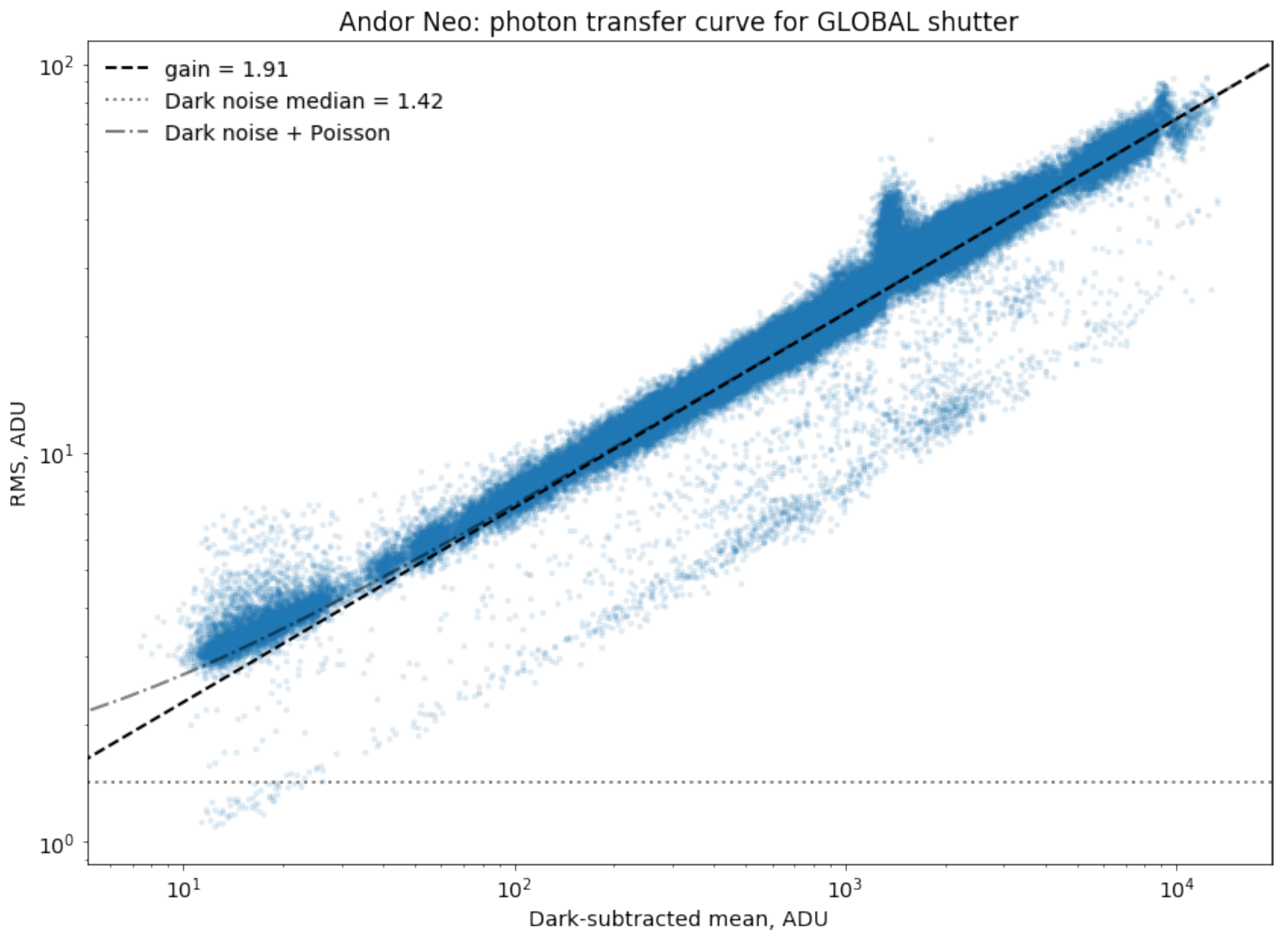}}
    \resizebox*{0.5\columnwidth}{!}{\includegraphics[angle=0]{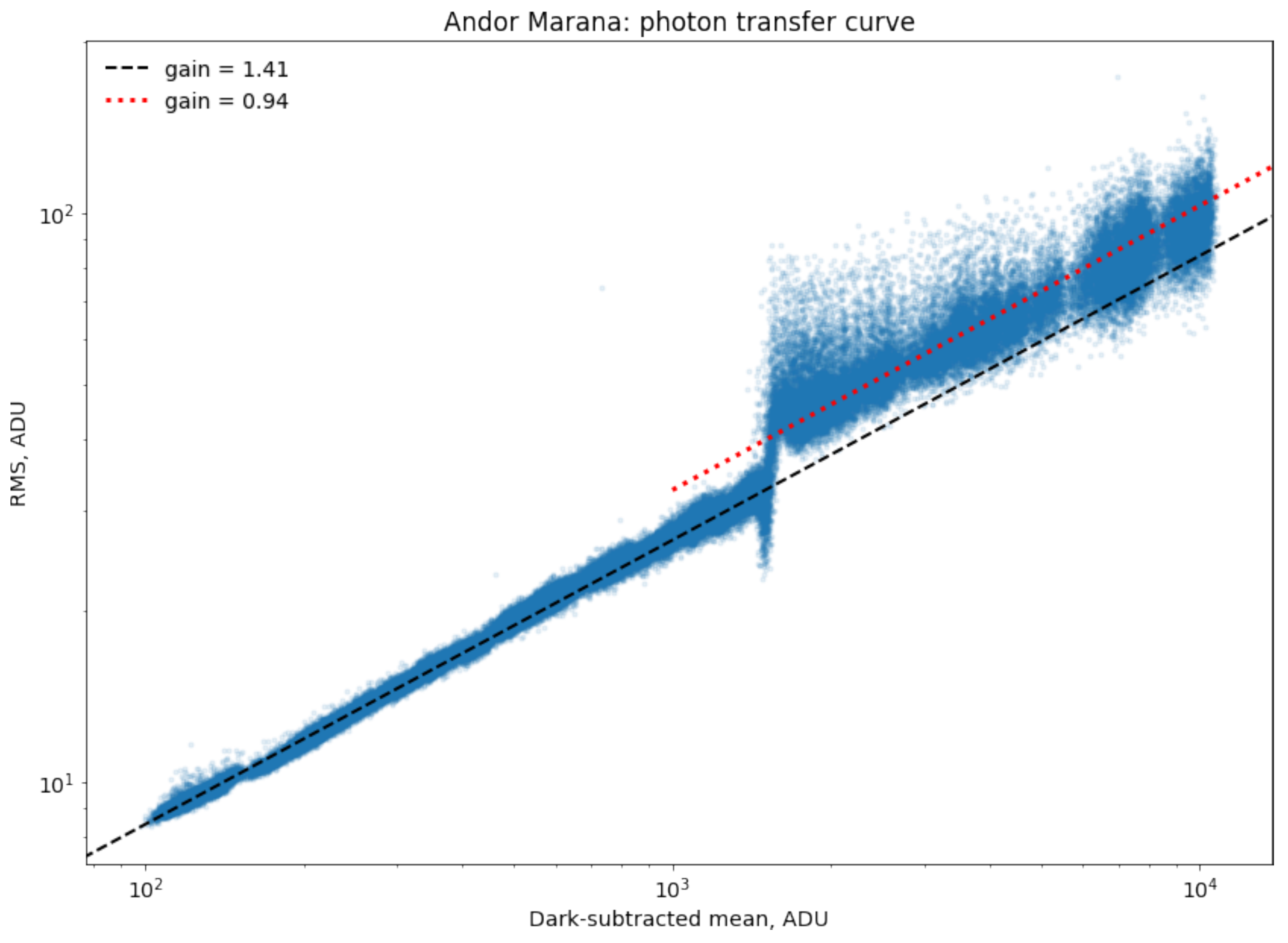}}
  }
  \caption{Photon transfer curve, i.e. dependence of a pixel temporal variance on (dark level subtracted) mean value
    for Andor Neo (left panel) and Andor Marana (right panel). The secondary cloud of points below main one for Andor Marana corresponds to a set of blemished pixels having smaller variance. It is absent for Andor Marana due to much smaller fraction of such pixels there. The jump at around 1500 ADU represents the transition between low-gain and high-gain amplifiers, providing different effective gains and having different spatial structures.
    \label{fig_ptc}}
\end{figure}

\begin{figure}[t]
  \centerline{
    \resizebox*{0.5\columnwidth}{!}{\includegraphics[angle=0]{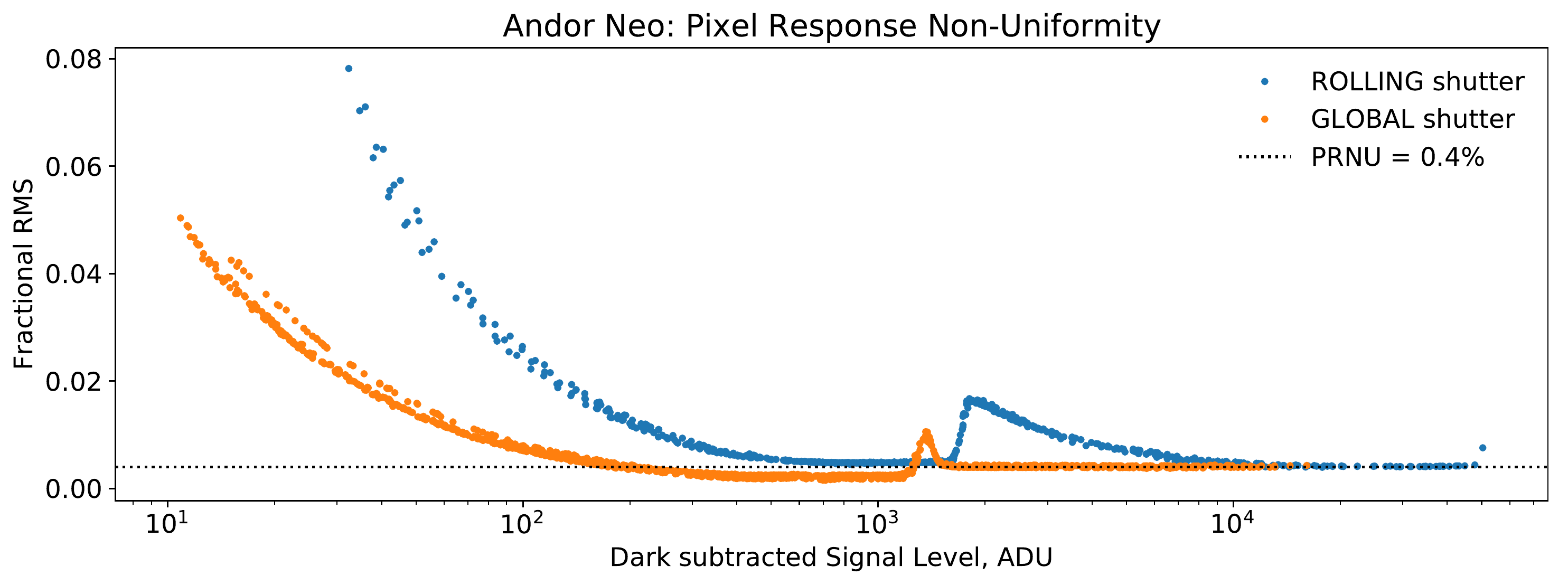}}
    \resizebox*{0.5\columnwidth}{!}{\includegraphics[angle=0]{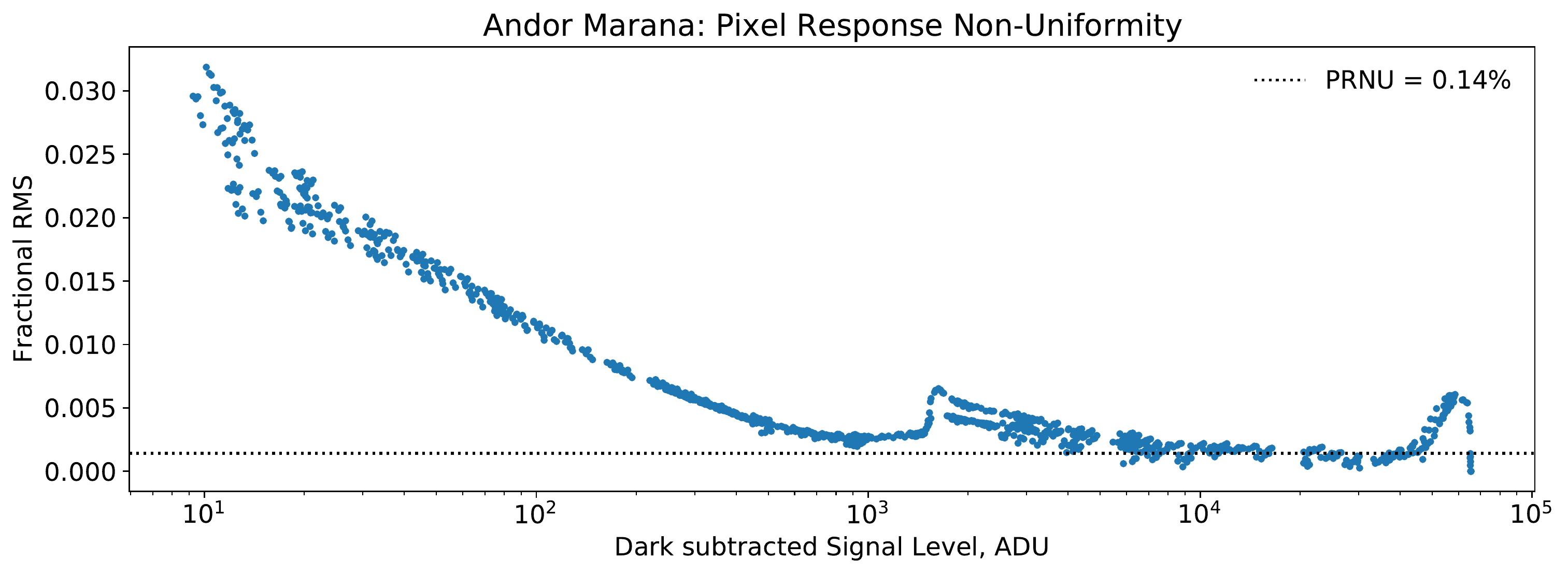}}
  }
  \centerline{
    \resizebox*{0.24\columnwidth}{!}{\includegraphics[angle=0]{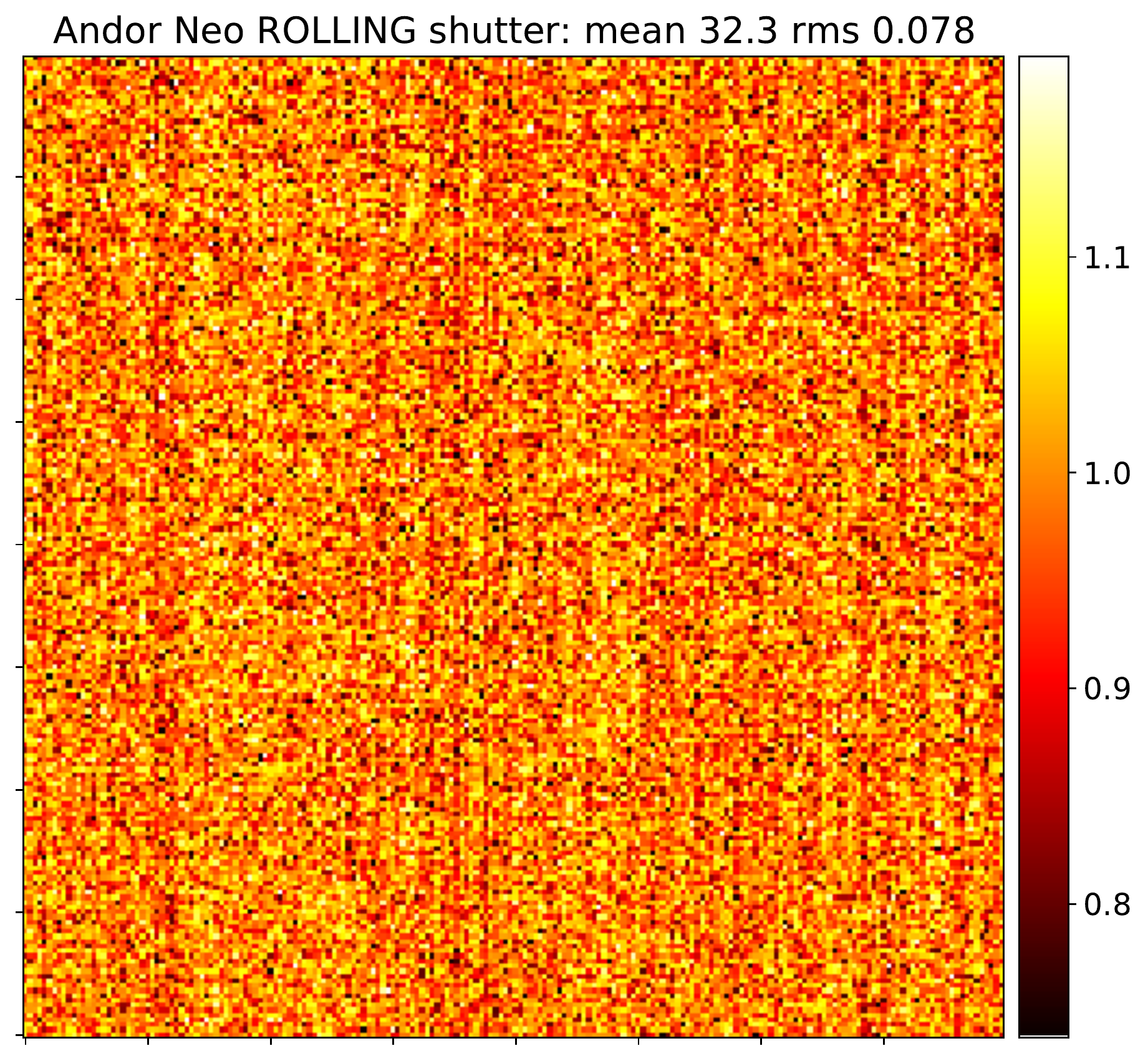}}
    \resizebox*{0.251\columnwidth}{!}{\includegraphics[angle=0]{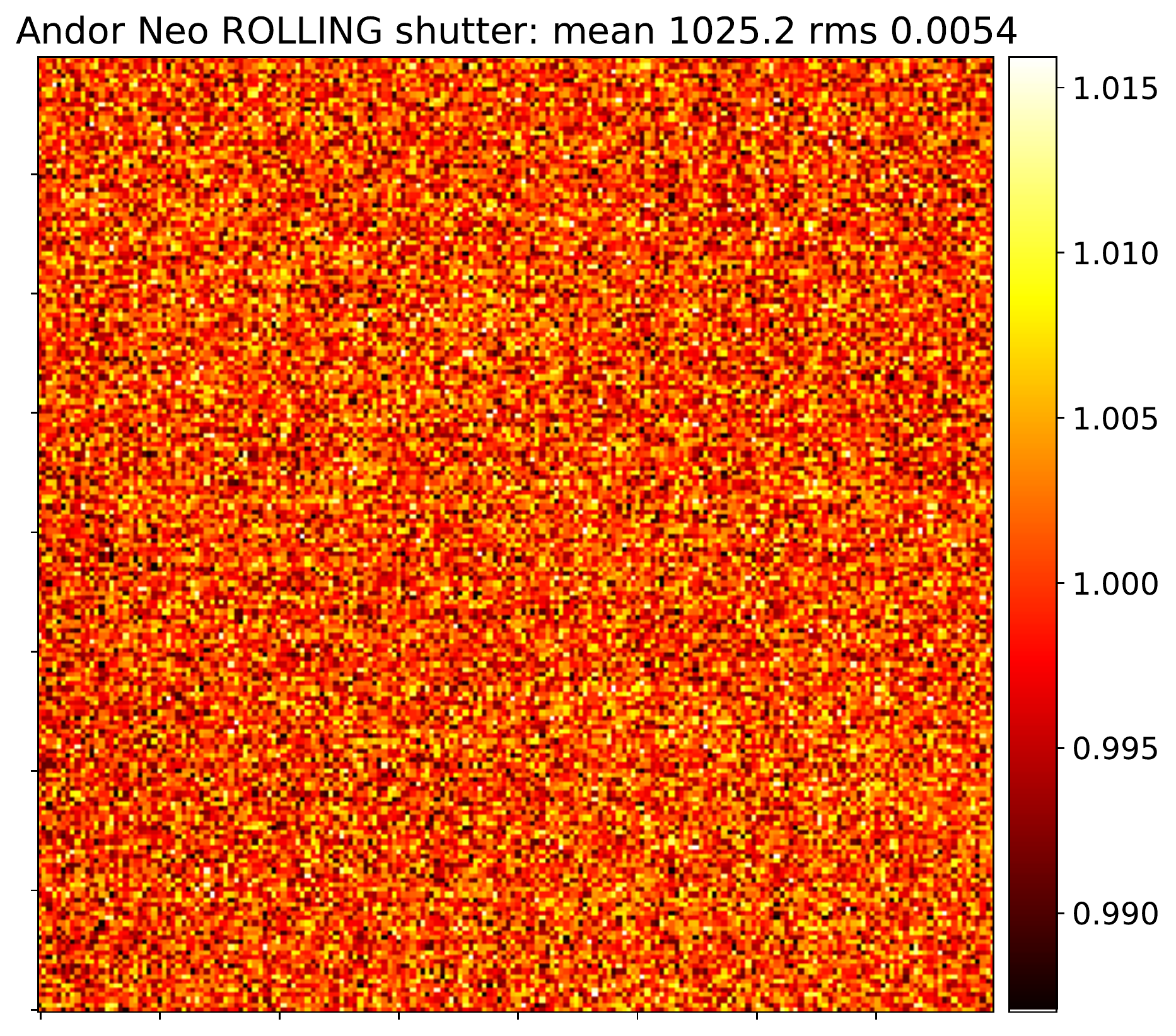}}
    \resizebox*{0.243\columnwidth}{!}{\includegraphics[angle=0]{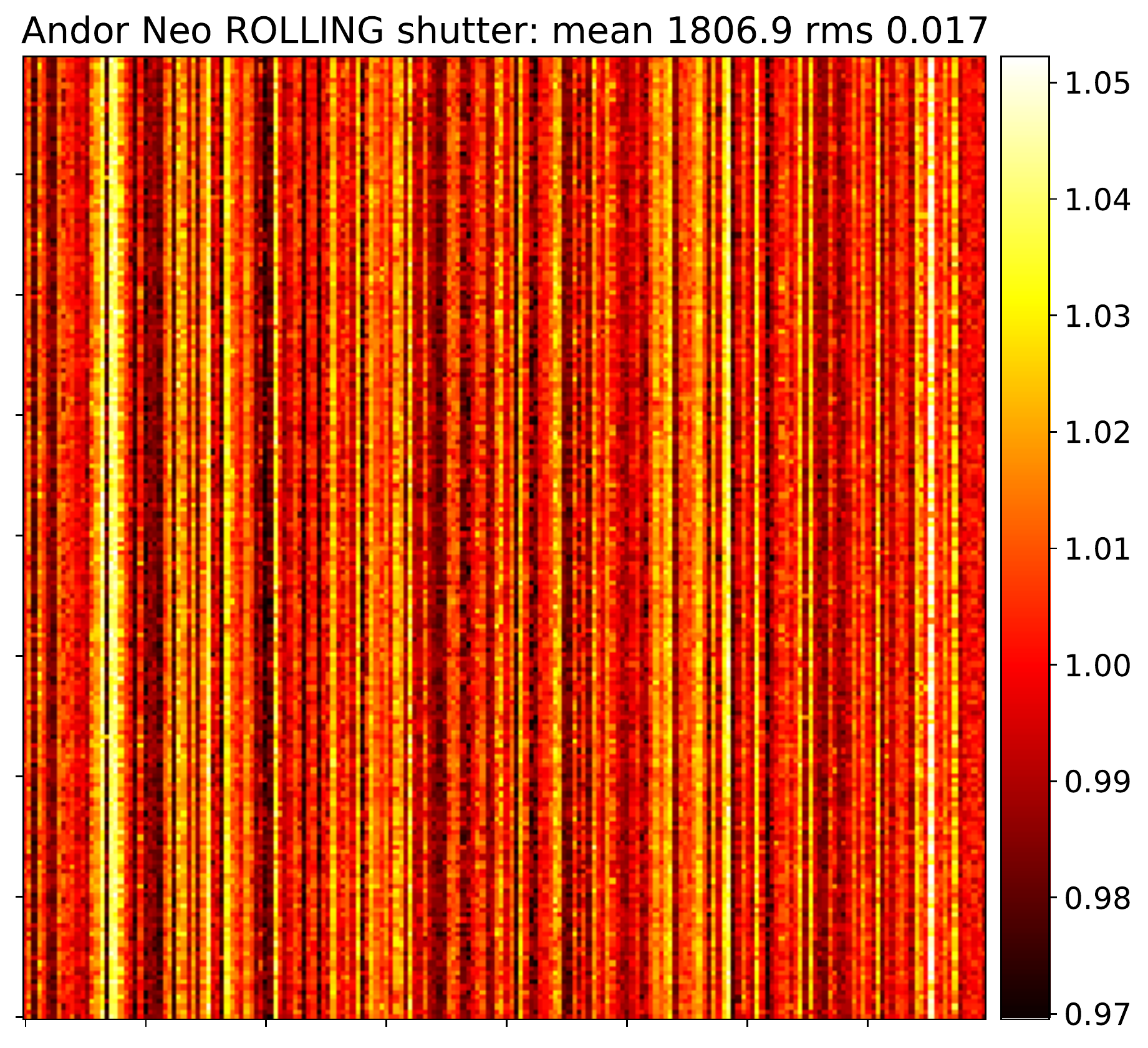}}
    \resizebox*{0.263\columnwidth}{!}{\includegraphics[angle=0]{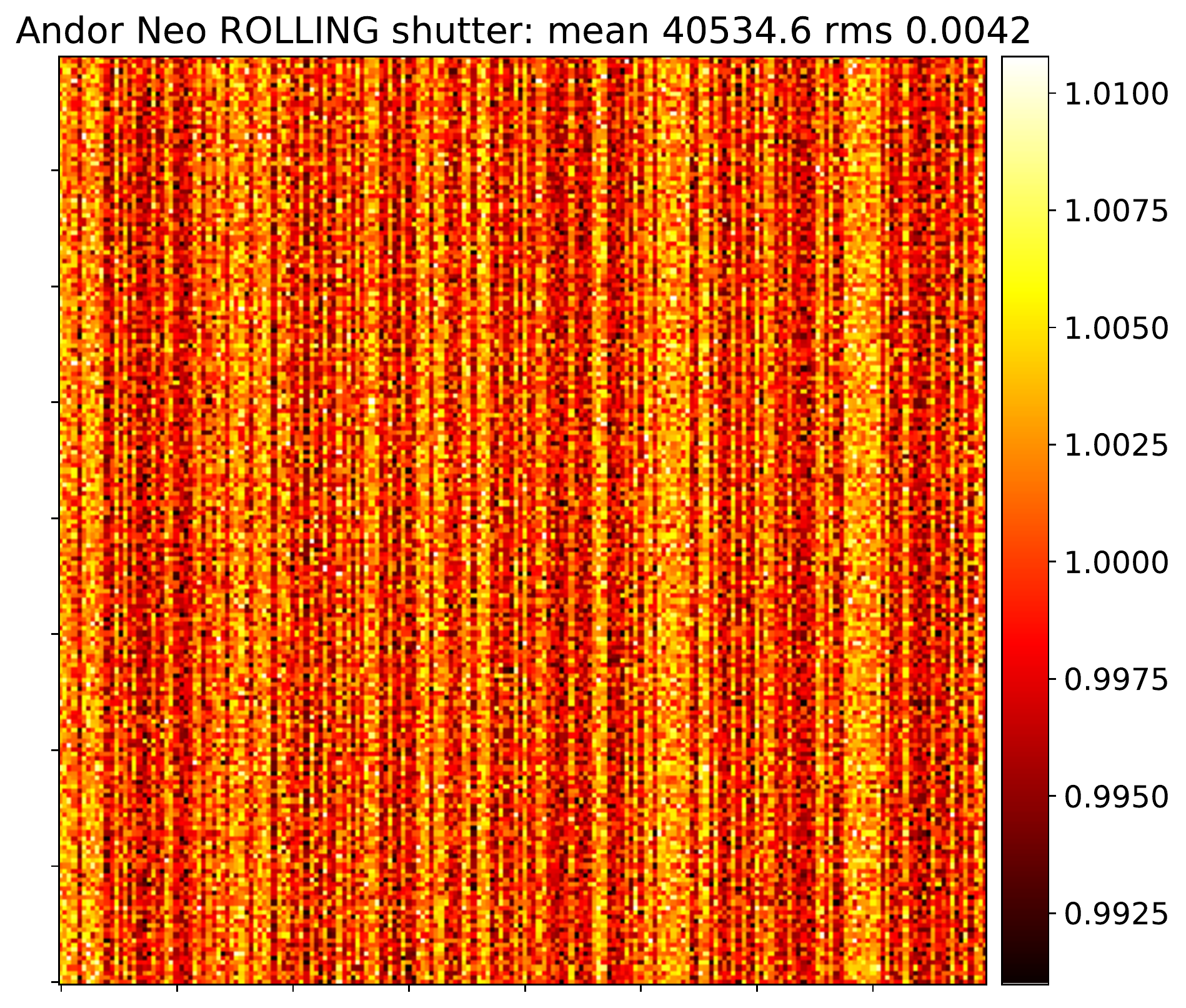}}
  }
  \caption{Pixel response non-uniformity (PRNU) after subtraction of an additive fixed pattern component (bias pattern) for different electronic shutter modes of Andor Neo (upper left panel) and for Andor Marana (upper right panel). Lower panels -- change of a small-scale structure of a flat field in the central part of a frame on Andor Neo with rolling shutter mode under different illumination levels. Prominent vertical striped structures of varying amplitudes are apparent both for lowest intensities and right after the transition to high gain amplifier at $\sim$1800 ADU. The amplitude of stripes then gradually decreases with intensity of illumination, but does not disappear completely up to saturation levels. Both the global shutter mode of Andor Neo and Andor Marana show the same behaviour, albeit with different amplitudes.
    \label{fig_prnu}}
\end{figure}


\begin{figure}[t]
  \centerline{
    \resizebox*{0.5\columnwidth}{!}{\includegraphics[angle=0]{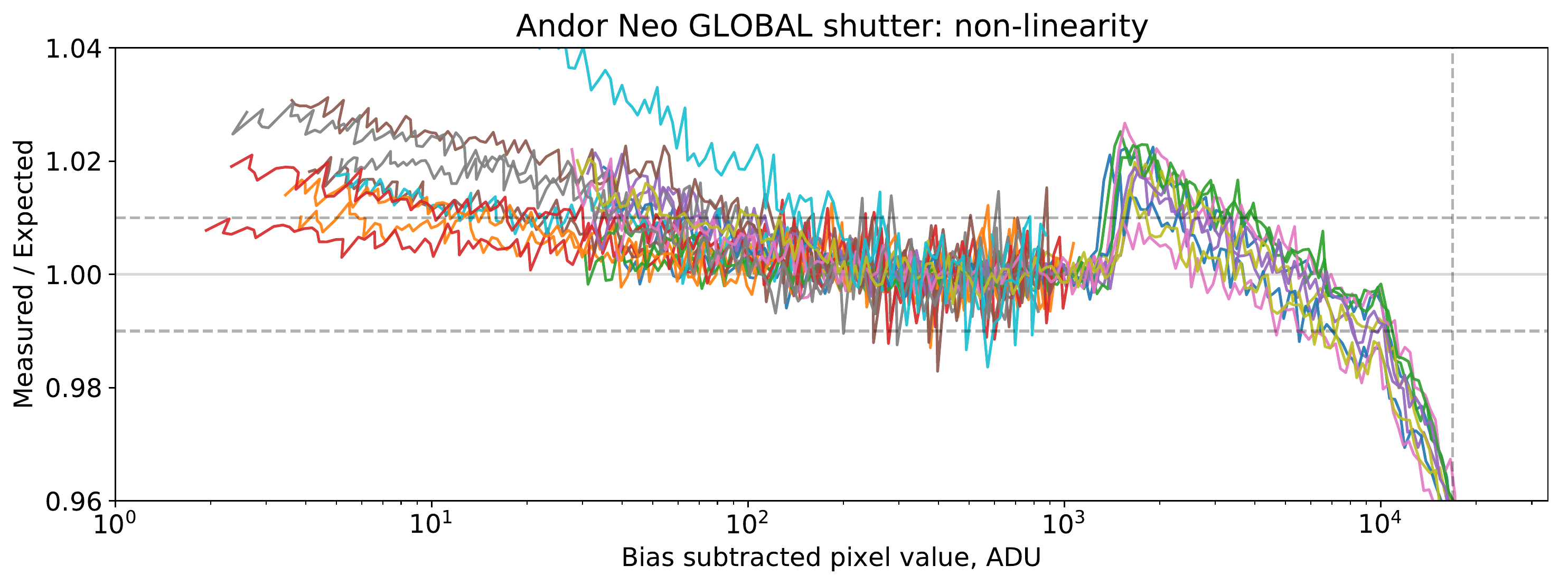}}
    \resizebox*{0.5\columnwidth}{!}{\includegraphics[angle=0]{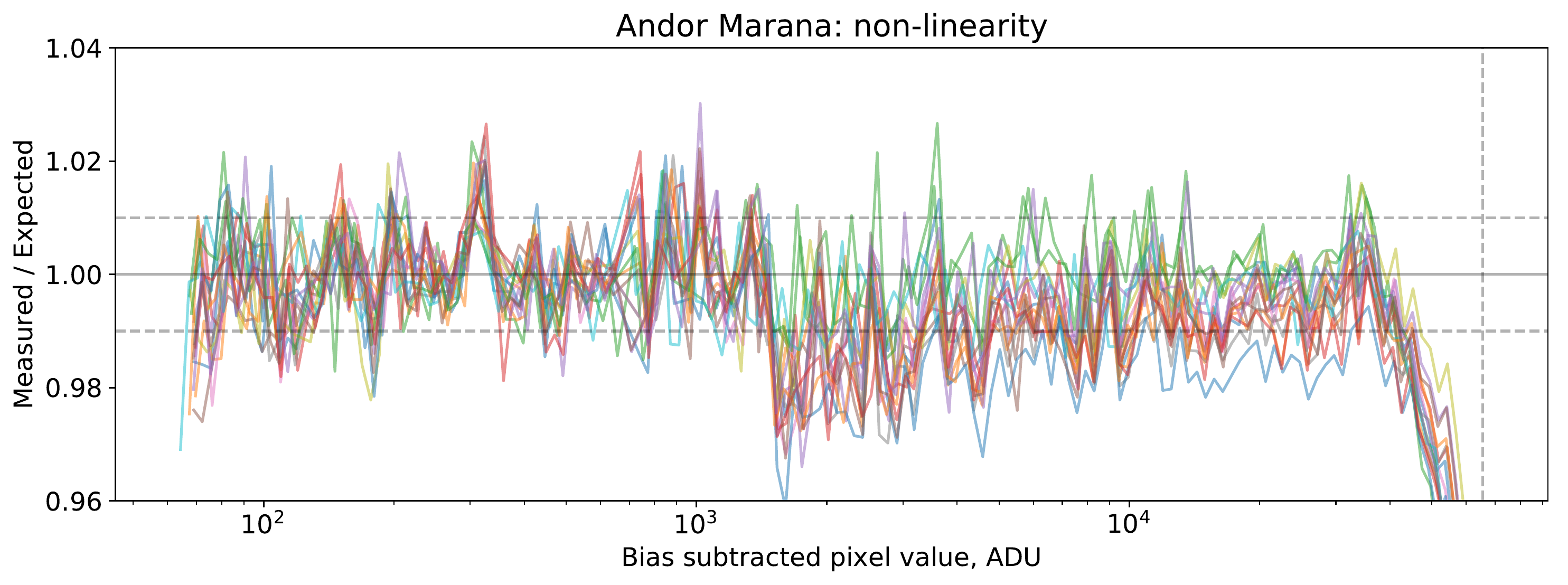}}
  }
  \caption{Linearity curves for a random set of pixels of Andor Neo (left panel) and Andor Marana (right panel) cameras. The curves represent the ratio of an actually measured signal to the one expected for a linear signal scaling with exposure time, with interval below 1000 ADU used to define a linear slope. Dashed vertical line to the right marks the position of a digital saturation (65535 ADU), systematic significant deviations from linearity start at about half of this value. The amplifier transition region is easily visible at around 1500 ADU, with the jump amplitude of typically less than couple of percents. Also, an additional change in the linearity curves is apparent at about half of saturation level for both cameras.
    \label{fig_linearity}}
\end{figure}

The photon transfer curve (PTC, the dependence of pixel RMS value on its mean, see \cite{ptbook}) for Marana camera shown in Figure~\ref{fig_ptc} displays a distinctive jump around 1500 ADU, corresponding to the transition between low-gain and high-gain amplifiers. The gain below the transition nicely corresponds to the one reported by manufacturer; above the transition, the effective gain drops by nearly two times, and its scatter across the pixels significantly increases. For Neo, the effective gains of values with lower and higher intensities are much better matched together. However, the spatial structures of pixel response -- pixel response non-uniformity, shown in Figure~\ref{fig_prnu} -- significantly changes with signal level both in amplitude and in shape, with distinctive column pattern appearing in flat fields after switching to high gain amplifiers. The amplitude of this pattern is sub-percent except for rolling gain mode of Neo camera, where it is up to couple of percents at the intensities close to amplifier transition.


This behaviour is equivalent to a slightly different signal response non-linearity between the pixels, which is illustrated by Figure~\ref{fig_linearity} which shows the ratio of measured signal in a pixel to the value expected if one assume purely linear response (e.g. signal directly proportional to exposure time under constant illumination). Indeed, signal response experiences sharp jump with couple of percents amplitude, and then a second feature at about half of saturation level. For Neo camera the overall non-linearity may reach up to 5\%\cite{karpov_2019}, while for Marana it seems a bit better.

The shape of non-linearity curves are quite similar between different pixels, but their exact amplitude and position of amplifier transition differ between them. So, in order to linearize the acquired image, one has to characterize every pixel of the camera independently (approximating it with e.g. piece-wise polynomial\cite{karpov_2019}).


Anyway, both non-linearity curves in Figure~\ref{fig_linearity} and pixel response non-uniformity plots of Figure~\ref{fig_prnu} suggest that the properties of flat fields are a bit different when acquired at different intensities, especially below and above the amplifier transition (i.e. 1500 ADU). Therefore, the procedure of flat field correction requires a special attention if one wants to reach sub-percent accuracy of measurements.

\section{Image persistence}

\begin{figure}[t]
  \centerline{
    \resizebox*{0.5\columnwidth}{!}{\includegraphics[angle=0]{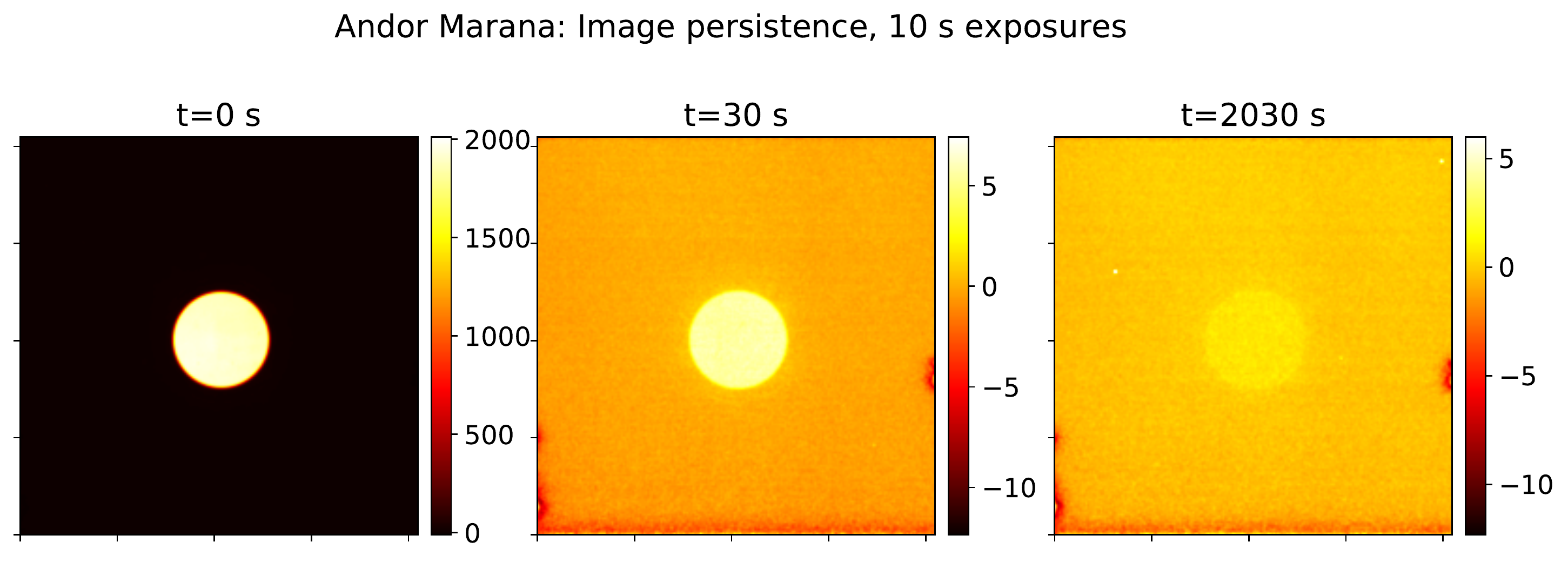}}
    \resizebox*{0.5\columnwidth}{!}{\includegraphics[angle=0]{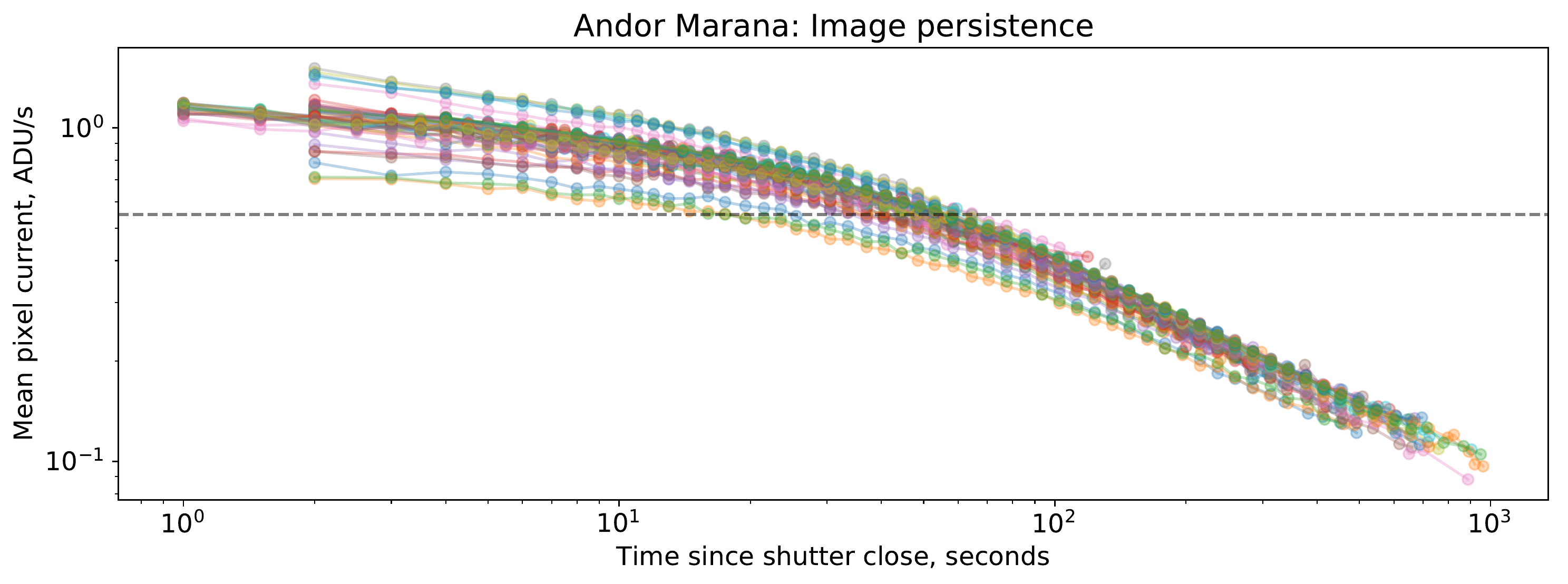}}
  }
  \centerline{
    \resizebox*{0.5\columnwidth}{!}{\includegraphics[angle=0]{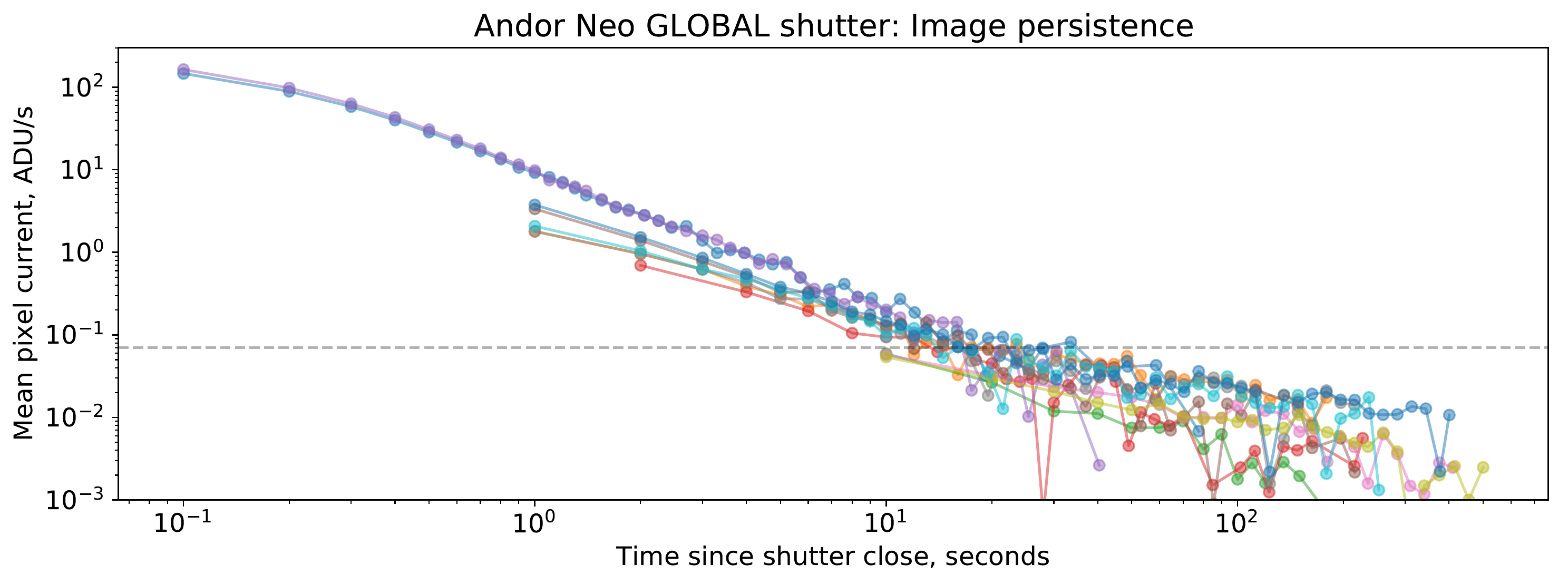}}
    \resizebox*{0.5\columnwidth}{!}{\includegraphics[angle=0]{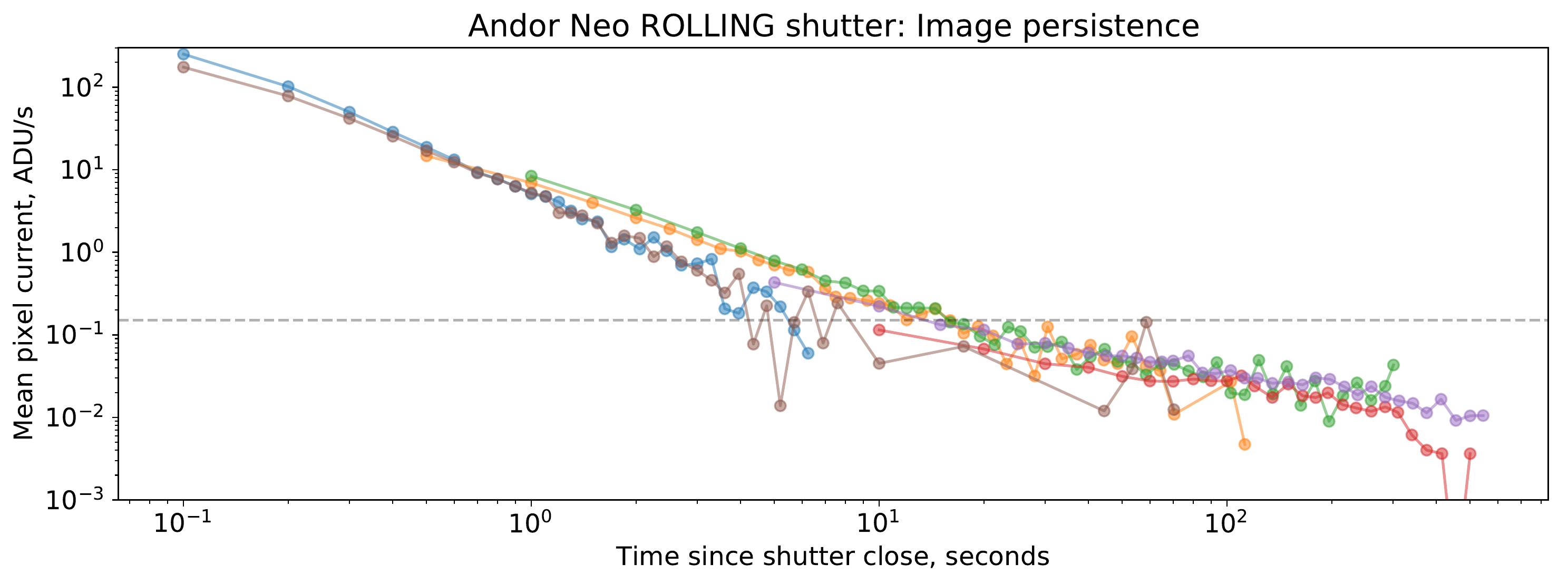}}
  }
  \caption{Image persistence characterization. Upper left panel -- full-frame images from Andor Marana right after bright spot illumination turned off and at two moments after that, with the camera constantly acquiring 10 s exposures. Even after 2030 s the residual image is still visible, though being well below dark current level. Upper right panel -- estimated dependence of residual image current (ADU per second per pixel) as a function of time since spot illumination is turned off. Lower panels -- the same for two different electronic shutter modes of Andor Neo.
    \label{fig_persistence}}
\end{figure}

Both Neo and Marana cameras display a prominent image persistence what may manifest for quite a long time, especially if the illumination pattern is extended (see upper left panel of Figure~\ref{fig_persistence} for an example image of a spot still visible on Marana frame more than half an hour after end of illumination).

This process is related to charge traps in the silicon substrate that capture the electrons when illumination level is high, and then gradually release them back over time. Therefore, the resulting signal manifests as an additional ``dark current'' with exponentially decaying amplitude. Our tests performed by means of illuminating the chip up to saturation and then measuring the residual current using sequences of frames with varying exposure times. They all give quite consistent results shown also in Figure~\ref{fig_persistence} -- the amplitude of residual current decays below the level of normal dark current in several tens of seconds on both cameras.

We were unable to detect the artifact residual images in our observations of stellar fields with exposures of tens of seconds, even for oversaturated stars and after rapid repointing of the cameras so that positions of the stars on the sensor change, which is correspond to the residual signal being well below other sources of noise on pixels scales. However, during routine observations of Mini-MegaTORTORA which have 10 frames per second frame rate and sometimes capture very bright flashes from rapidly moving satellites (due to rotation of their solar panels), a faint and rapidly fading afterglows lasting up to couple of seconds are eventually visible.

On the other hand, as neither cameras have a mechanical shutter, the parasitic illumination of the chip even when there is no readout may still cause a long lasting extended features in the images acquired later.

\section{On-sky testing}\label{sec_sky}

\begin{figure}[t]
  \centerline{
    \resizebox*{0.542\columnwidth}{!}{\includegraphics[angle=0]{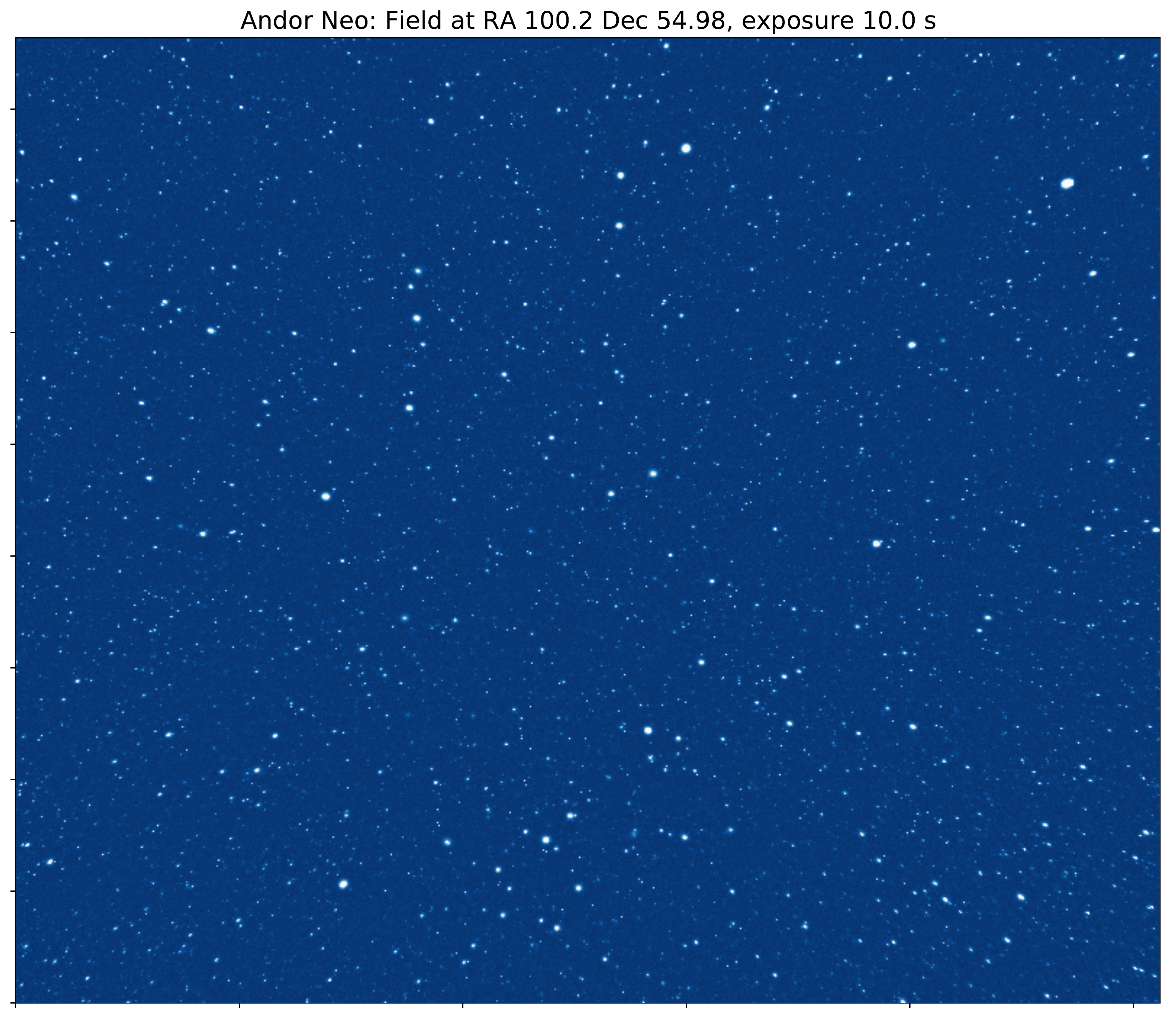}}
    \resizebox*{0.458\columnwidth}{!}{\includegraphics[angle=0]{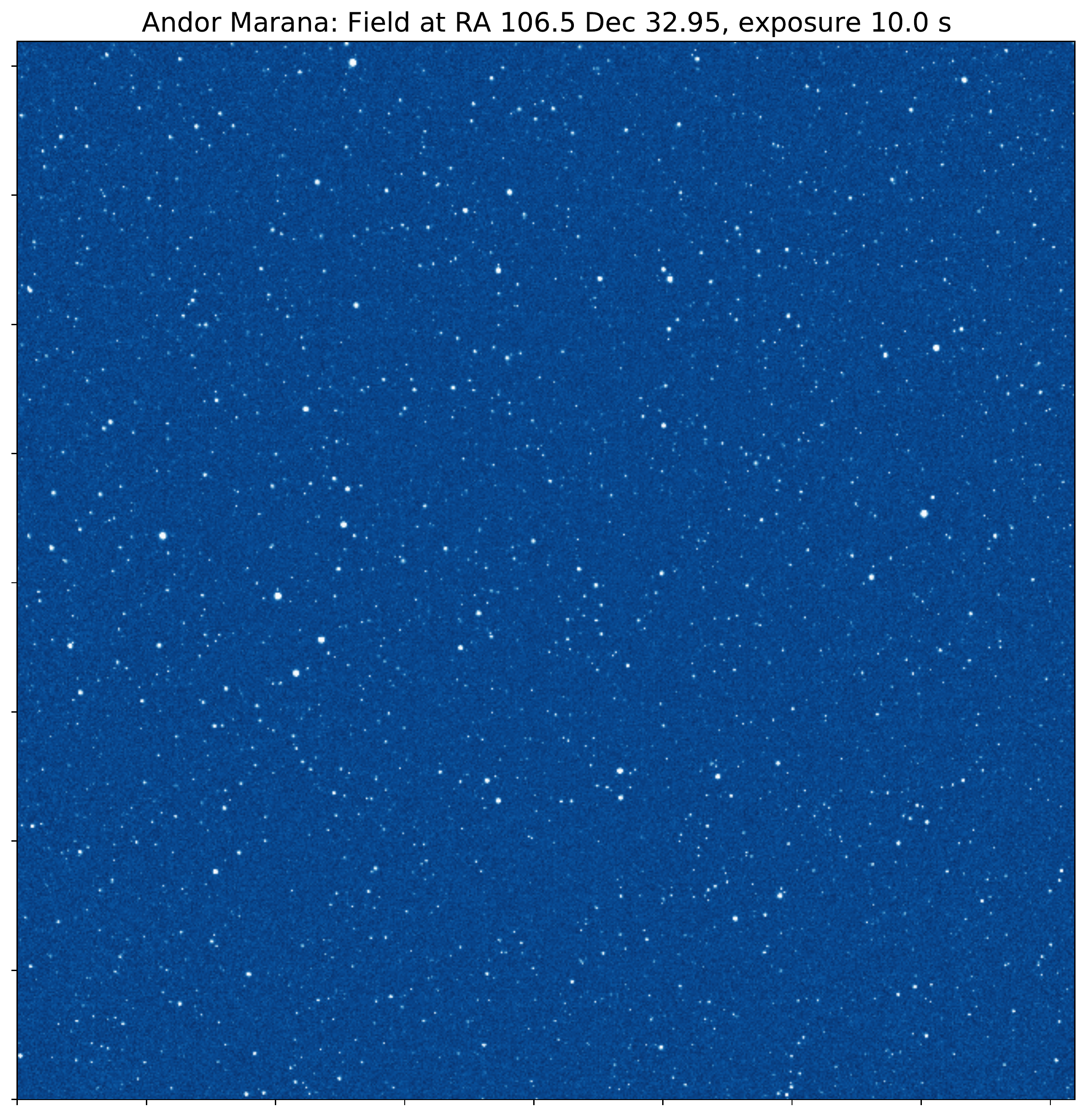}}
  }
  \caption{Single sky frames acquired during on-sky testing experiments with Andor Neo (left panel) and Andor Marana (right panel) cameras. Andor Neo camera used is a part of Mini-MegaTORTORA system\cite{beskin_astbull,karpov_2019}, and thus is equipped with Canon EF85 f/1.2 fast lens with no photometric filters installed. Field of view is 11.3$^{\circ}$x9.5$^{\circ}$ with 15.9$''$/pixel scale and 3.2 pixels median FWHM mostly defined by the large PSF wings formed by the objective lens. Andor Marana camera is equipped with Nikkor 300 f/2.8 lens with no color filters, giving field of view size of 4.26$^{\circ}$x4.26$^{\circ}$ with 7.5$''$/pixel scale. Median FWHM of the stars is 2.1 pixels, making the image nearly critically sampled. Both frames are dark subtracted and flat-fielded using sky flats. Note the absence of cosmetic defects typical for CCD frames -- hot and dark columns, charge bleedings from oversaturated stars, etc.
    \label{fig_sky}}
\end{figure}


\begin{figure}[t]
  \centerline{
    \resizebox*{0.5\columnwidth}{!}{\includegraphics[angle=0]{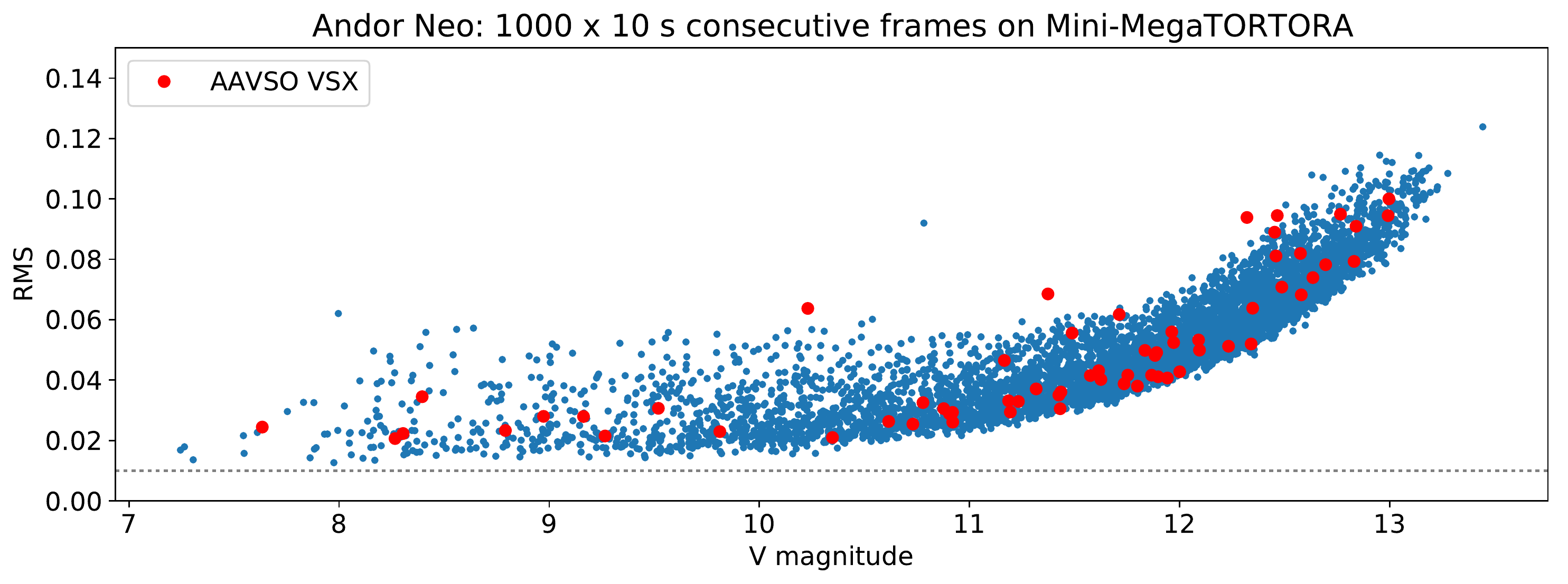}}
    \resizebox*{0.5\columnwidth}{!}{\includegraphics[angle=0]{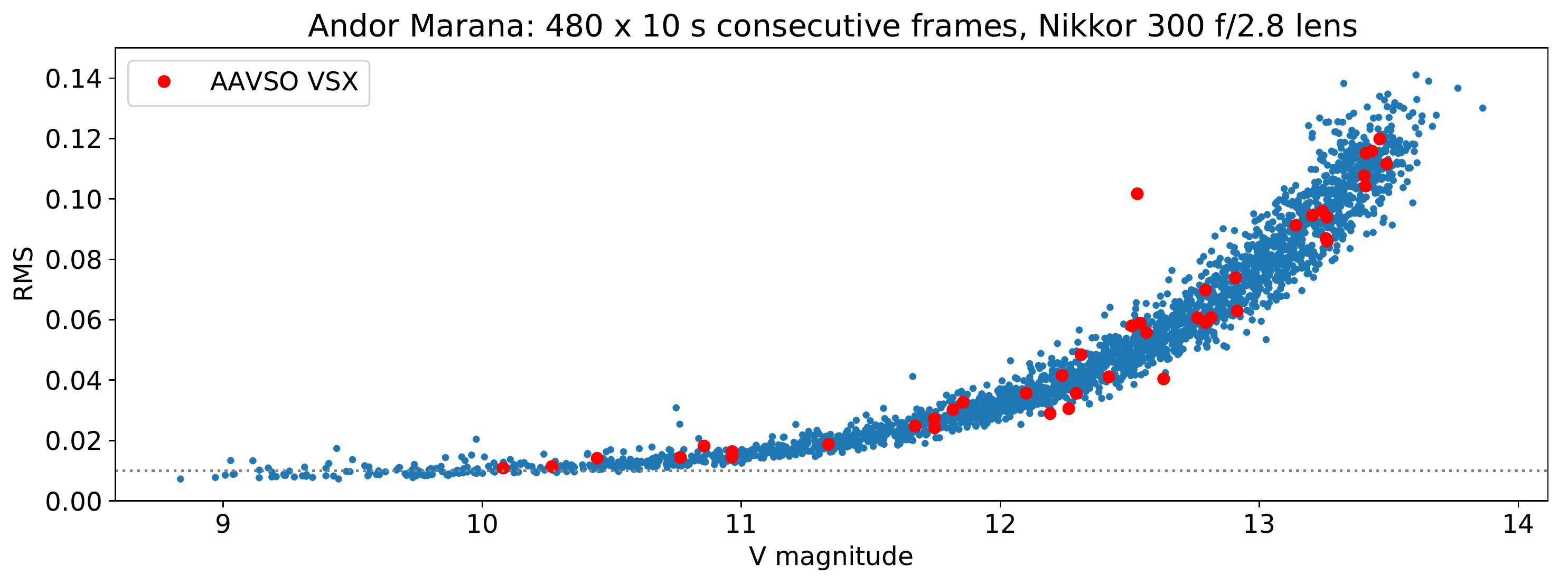}}
  }

  \centerline{
    \resizebox*{0.44\columnwidth}{!}{\includegraphics[angle=0]{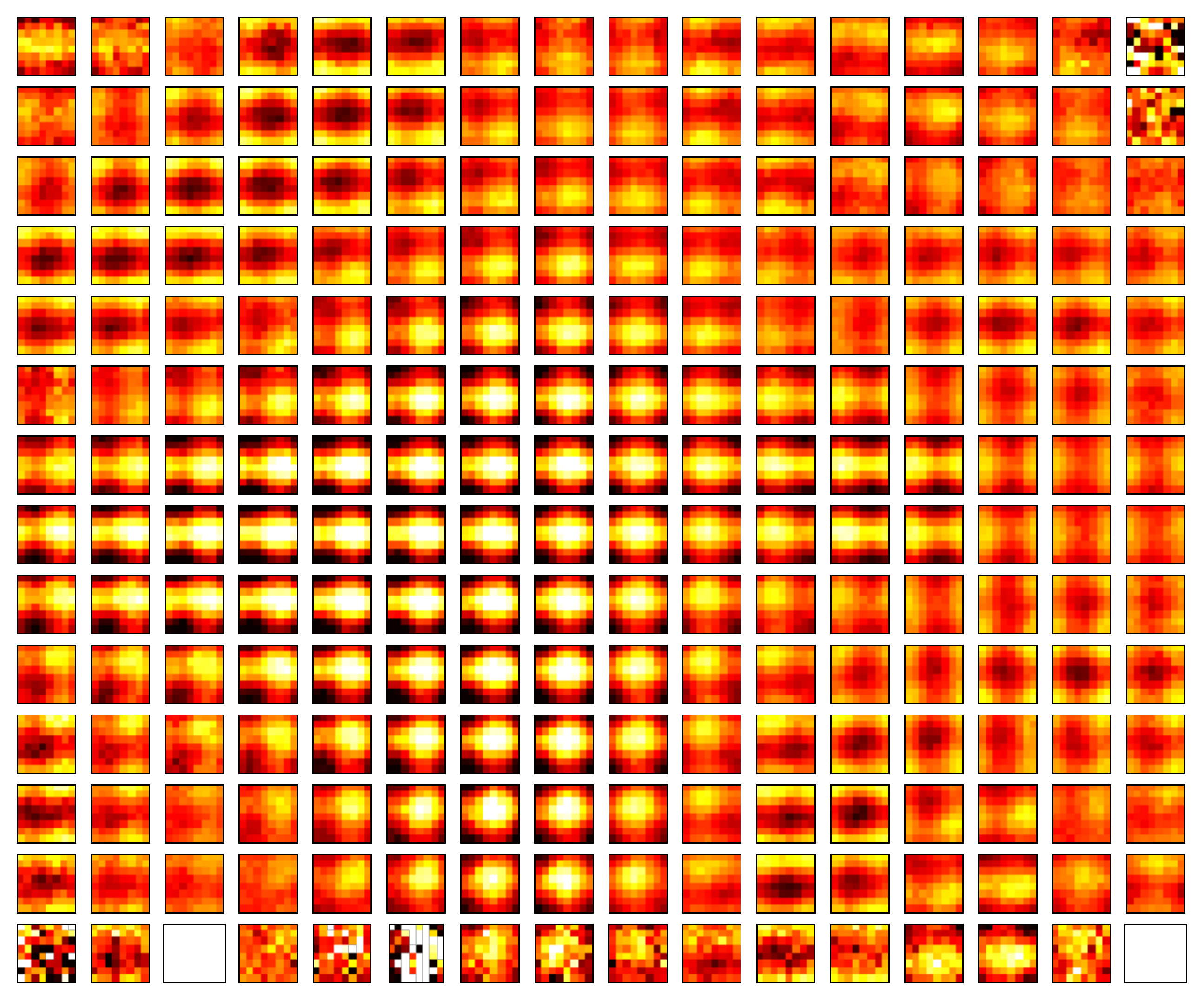}}
    \resizebox*{0.0472\columnwidth}{!}{\includegraphics[angle=0]{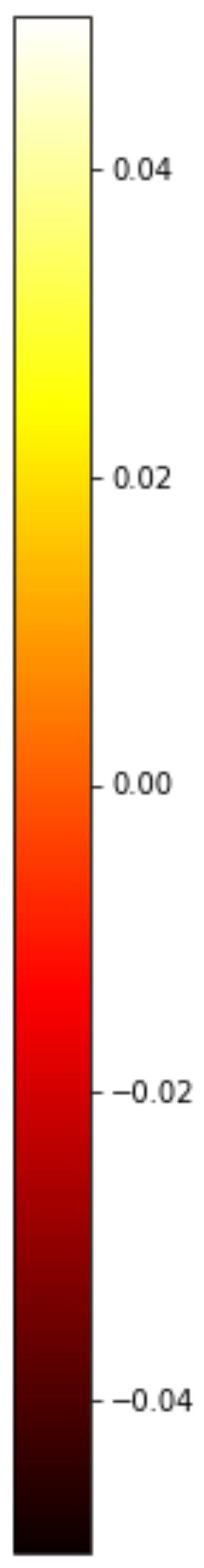}}

    \resizebox*{0.51\columnwidth}{!}{
      \begin{tabular}[b]{l}
        \includegraphics[angle=0]{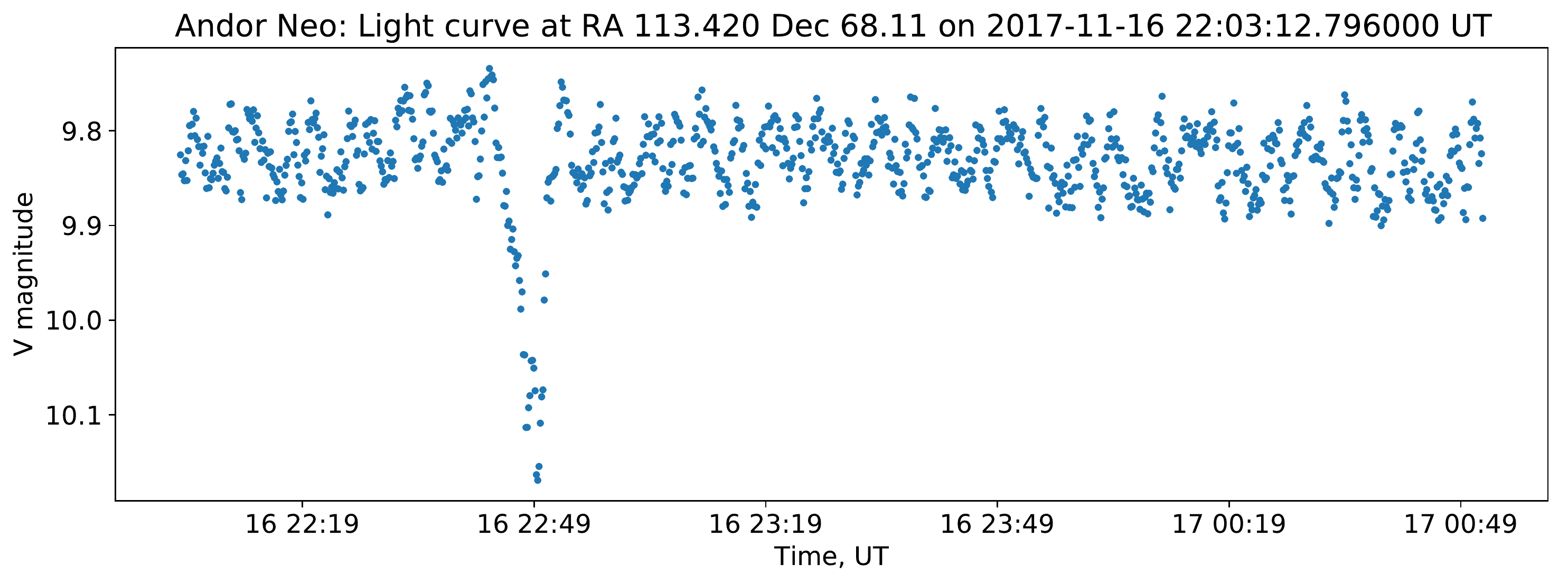}\\
        \includegraphics[angle=0]{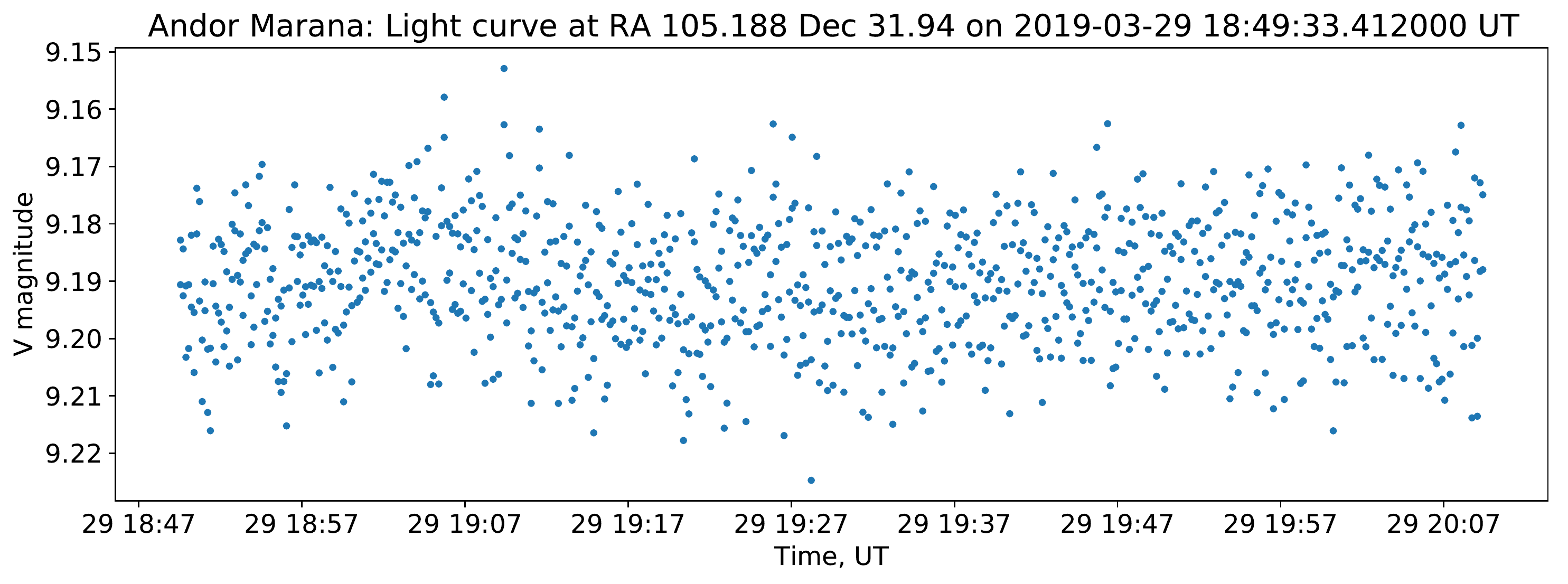}
      \end{tabular}%
    }
  }
  \caption{On-sky performance of the cameras. Upper panels -- the scatter of photometric measurements of individual star along the sequence of sky images (acquired on an imperfect mounts causing stars to slowly drift across the detector pixels) versus its mean value. Red circles represent known variable stars from AAVSO VSX database. For Andor Neo, operated together with Canon EF85 f/1.2 fast lens of a Mini-MegaTORTORA system, the plot shows a significant additional scatter due to sub-pixel sensitivity variations caused by microlens raster on top of the chip. Andor Marana, operated with Nikkor 300 f/2.8 lens and lacking microlens raster due to back-illuminated chip design, shows no additional scatter and much more stable photometry.
    Lower left panel -- the map of sub-pixel sensitivity variations for Andor Neo as used with Canon EF85 f/1.2 fast lens of a Mini-MegaTORTORA system, where each sub-image represent the difference of measured stellar magnitude as a function of a sub-pixel peak position at different places of a chip. Color-coding represent the 5\% amplitude of variations. Middle right panel -- the light curve of a single star slowly drifting across Andor Neo pixels in the same setup. Lower-amplitude oscillations correspond to sub-pixel sensitivity variations from lower left panel, while more prominent dip -- to the passage of photometric aperture over a blemished pixel. Both of these effects are missing for a light curves on Andor Marana, example of which is shown in lower right panel.
    \label{fig_sky_rms}}
\end{figure}

On-sky testing of the cameras consisted of a series of continuous observations of a fixed sky positions with moderate exposures in order to assess the photometric performance and achievable stability of the data. The acquired frames (examples for both cameras are shown in Figure~\ref{fig_sky}) were dark subtracted and flattened using standard operational flat field images of Mini-MegaTORTORA system (for Neo camera) or evening flats constructed by a median averaging of evening sky images (for Marana camera). Then every frame was astrometrically calibrated using {\sc Astrometry.Net} \cite{astrometry.net} code. Then the forced aperture photometry with a small (2 pixels radius) apertures was performed using the routines available in {\sc SEP} \cite{SEP} Python package (based on original SExtractor code by \cite{sextractor}) at the positions of all sufficiently bright stars from Gaia DR2 catalogue\cite{gaiadr2} that fit into the frame. This way we mostly avoided the additional photometric noise due to centroiding errors which may be sufficient for a nearly critically sampled stellar profiles. Then, on every frame the zero point model was constructed by fitting the measured brightness with Gaia DR2 photometric data (G magnitudes and BP-RP colors what we statistically re-calibrated to Johnson-Cousins system by cross-matching them with Stetson standard stars) and a model including fourth-order spatial polynomial to to compensate imperfections of evening flats as well as positional-dependent aperture correction due to changes of stellar PSF, and a color term to take into account the (time dependent) difference between our unfiltered photometric system and a Johnson-Cousins one. This way we may estimate the $V$ magnitude of stars visible on the frame assuming their colors correspond to the catalogue values (or just fit for them independently, if the observations span long enough time interval with changing atmospheric conditions\cite{karpov_photometric_calibration}).

The stability of photometric measurements over long sequences of consecutive frames is illustrated by a standard ``scatter vs magnitude'' plots shown in upper panels of Figure~\ref{fig_sky_rms}. While Marana camera provides sufficiently stable measurements, reaching better than 1\% precision for brighter stars which is consistent with expected Poissionian statistics, the photometry from Neo camera shows significant additional scatter with amplitude of several percents. Its detailed investigation reveals that this scatter has a systematic nature correlated with a stellar centroid position within the pixel (which is slowly changing over time due to imperfect mount tracking), with exact shape and amplitude of this dependence slowly varying across the field of view but reaching $\pm$5\% in the frame center (see lower left panel of Figure~\ref{fig_sky_rms}). This is caused by an interplay of a fast (f/1.2) objective lens with the microlens raster array covering the pixels of the CIS2521 chip of Neo camera in order to increase its quantum efficiency. Another source of additional noise in Neo data is the emergence of blemished pixels within the photometric aperture. Middle right panel of Figure~\ref{fig_sky_rms} shows a typical intensity dip related with the drift of a star across such blemished pixel, which is being interpolated by camera onboard FPGA from the readings of pixels surrounding it. The same plot also shows a quasi-periodic oscillations of intensity due to intra-pixel sensitivity variations. Both of these effects are not visible in the data from Marana camera, which is consistent with much smaller number of blemished pixels there, and its chip being back-illuminated with no microlens raster array on top of it (and additionally a slower focal ratio of the objective we used for the tests).

\section{Conclusions}\label{sec_conclusions}

The comparison of two models of Andor cameras based on different generations of sCMOS chips from different manufacturers according to their specifications is shown in Table~\ref{tab_marana}. Our results confirms that Marana dark current is indeed significantly larger, especially with on-board glow correction disabled.

The linearity of Marana is nearly perfect up to approximately half of saturation level, with just an occasional jump of typically less than 2\% at the amplifier transition region. The effective gain, however, drops  by nearly a two times there, also changing the spatial properties of flat fields above approximately 1500 ADU by revealing the structure of column level amplifiers. The behaviour of Neo camera is a bit worse, with more complex systematic trends in linearity, again spanning couple of percents level both on very low intensities of light and above the level of transition to high gain amplifiers. On the other hand, the gain values on lower and higher intensities are much better matched on Neo.

The amount of blemished pixels in Marana is nearly 30 times smaller than in Neo, where they occupied up to 0.66\% of all pixel area and significantly affected the photometric performance. The noise at the dark level shows a random telegraph signal features in a small (0.01\%) with a jumps amplitude around 10 ADU. Moreover, the intensity region around amplifier transitions also shows a pixel value excess jumps related to switching between readings of low gain and high gain amplifiers. All other intensity regions show quite stable and mostly normally distributed pixel readings over time. On lower intensities, the noise shows some spatial correlations related to onboard overscan subtraction in Neo cameras and to some other effects in Marana, both fading away as intensity of incoming light increases. The pixel response non-uniformity also decreases towards sub-percent levels, but the small-scale structure of flat fields drastically changes with transition between low gain and high gain amplifiers, with distinct column structure apparent in the latter (high intensity) regime on both cameras. This effect may impact the flat field correction on sub-percent level.

Our on-sky tests of Marana display promising performance, with a photometric precision easily reaching 1\% on a sequence of consecutive frames even with critically sampled (FWHM$\approx$2 pixels) stellar profiles, and without any signs of sub-pixel position related systematics like the ones evident in Neo, which employs a microlens raster in front of a sensitive part of every pixel.

Moreover, the overall quality of the images are nice and free of typical CCD cosmetic problems -- hot and dark columns, bleedings from oversaturated stars, etc -- due to per-pixel read-out, which significantly increases the amount of scientifically usable pixels over the sensor area. 

Both cameras display a moderate image persistence, but according to our tests it does not affect typical observations of point sources like stellar objects.

That all said, we may safely conclude that Andor Marana sCMOS is indeed a very promising camera for a sky survey applications, especially requiring high temporal resolution, and exceeds Andor Neo in nearly all aspects of it.

\acknowledgments 

This work was supported by European Structural and Investment Fund and the Czech Ministry of Education, Youth and Sports (Project CoGraDS -- CZ.02.1.01/0.0/0.0/15 003/0000437). Authors are grateful to Andor, Oxford Instruments Company for providing the Marana camera used for testing. Mini-MegaTORTORA system and Neo cameras installed in it belong to Kazan Federal University, and the study is partially performed according to the Russian Government Program of Competitive Growth of Kazan Federal University.

\bibliography{cmos} 
\bibliographystyle{spiebib} 

\end{document}